\documentclass[showpacs,pra,superscriptaddress,onecolumn]{revtex4}
\usepackage{amsmath}
\usepackage{graphicx}
\usepackage{subfigure}

\begin{document}

\title{Transfer and scattering of wave packets by a nonlinear trap}
\author{Kai Li}
\affiliation{Department of of Mathematics and Statistics, University of Massachusetts,\\
Amherst, MA 01003-9305, USA}
\author{P. G. Kevrekidis}
\affiliation{Department of of Mathematics and Statistics, University of Massachusetts,\\
Amherst, MA 01003-9305, USA}
\author{Boris A. Malomed}
\affiliation{ICFO-Institut de Ciencies Fotoniques,\\
Mediterranean Technology Park, 08860 Castelldefels (Barcelona), Spain\\
Department of Physical Electronics, School of Electrical Engineering,
Faculty of Engineering, Tel Aviv University, Tel Aviv 69978, Israel\thanks{%
permanent address}}
\author{D.J.\ Frantzeskakis}
\affiliation{Department of Physics, University of Athens, Panepistimiopolis, Zografos,
Athens 157 84, Greece}

\begin{abstract}
In the framework of a one-dimensional model with a tightly localized self-attractive
nonlinearity, %
we study the
formation and transfer (dragging) of a trapped mode by ``nonlinear
tweezers", as well as the scattering of coherent linear wave packets on the
stationary localized nonlinearity. The use of the nonlinear trap for the
dragging allows one to pick up and transfer the relevant structures
without
grabbing surrounding ``garbage". %
A stability border for the dragged modes is identified by means of of
analytical estimates and systematic simulations. In the framework of the
scattering problem, the shares of trapped, reflected, and transmitted wave
fields are found. Quasi-Airy stationary modes with a divergent norm, that
may be dragged by the nonlinear trap moving at a constant acceleration, are
briefly considered too.
\end{abstract}

\maketitle

\section{Introduction}

Solitons, in the form of robust localized pulses propagating in nonlinear
dispersive media, have been studied extensively in various physical contexts
\cite{dau}. Many works have been dealing with the soliton dynamics in media
featuring uniform nonlinearities, modeled by wave equations with constant
nonlinearity coefficients. More recently, much attention has been
drawn to the study of solitons in structured media, which feature spatial
modulations of the local nonlinearity strength (see recent review \cite%
{Barcelona}), including the case when stable \emph{bright} solitons may be
supported by \emph{self-defocusing} nonlinearity whose strength grows
towards the periphery \cite{Barcelona2}. In solid-state physics, effective
potentials induced by the nonlinearity modulation are often called \textit{%
pseudopotentials} \cite{Harrison}. Recently developed experimental
techniques make it possible to realize such structures in photonics and
Bose-Einstein condensates (BECs). Particularly, in photonic media, the
modulation profiles can be induced by nonuniform distributions of resonant
dopants affecting the local nonlinearity, which may be created by means of
in-diffusion technology \cite{technology}. Another possibility is to use
photonic crystals with voids infiltrated by liquids, which are matched to
the host material in terms of the refractive index, but feature a mismatch
in the Kerr coefficient \cite{infiltration}. On the other hand, in the
context of BECs, the local nonlinearity can be spatially modulated by means
of the Feshbach resonance, induced by nonuniform magnetic \cite{Inouye} or
electric \cite{Marinescu} fields, as well as by an appropriate optical field
\cite{Fedichev}.

One-dimensional (1D) solitons were studied theoretically in the context of
pseudopotential structures in BECs \cite{BECpseudopotential,Nir} and optics
\cite{Hao,Barcelona2}. The possible use of 1D nonlinearity-modulation
profiles was also elaborated in models of matter-wave lasers, designed so as
to release periodic arrays of coherent pulses \cite{laser,Humberto}
(although the first experimentally realized prototypes of atomic-wave lasers
were built in a different way \cite{real-laser}). Self-focusing
pseudopotentials were theoretically studied in the two-dimensional (2D)
geometry too, demonstrating that it is much more difficult (but possible) to
stabilize 2D solitons in such settings, than using familiar linear
potentials or lattices \cite{Sakaguchi2,Fibich,JMO}, while the
above-mentioned self-defocusing structures solves this problem easily \cite%
{Barcelona2}.

The simplest example of the pseudopotential corresponds to the nonlinearity
concentrated at a single point, which is accounted for by the $\delta $%
-function \cite{Azbel}. In optics, it may represent a planar waveguide with
a narrow nonlinear stripe embedded into it (stable spatial solitons
supported by narrow stripes carrying the quadratic, rather than cubic,
nonlinearity were found in Refs.~\cite{Canberra,Asia}). In BECs, the
localized nonlinearity may be imposed by a tightly focused laser beam
inducing the Feshbach resonance. A relevant extension of such a nonlinear
spot is the model of the \textit{double-well} pseudopotential; in the
simplest form, it is based on a symmetric set of two $\delta $-functions
multiplying the cubic term in the respective nonlinear Schr\"{o}dinger (NLS)
equation \cite{DeltaSSB}, or the quadratic nonlinearity in the model of the
second-harmonic generation \cite{Asia}. In either case, full analytical
solutions have been found for symmetric, asymmetric and antisymmetric
localized modes trapped by the set of the two $\delta $-functions. In
addition to that, full symmetric and asymmetric solutions can be found \cite%
{Valera} for a similar discrete model, with two cubic-nonlinear sites
embedded into a linear host chain, which was introduced (without the
consideration of the spontaneous symmetry breaking) in Refs. \cite{Tsironis}%
. The symmetry-breaking problem was also recently analyzed in the
two-dimensional setting, for a symmetric pair of nonlinear circles embedded
into a linear host medium \cite{JMO}.

The model with the single nonlinear $\delta $-function readily gives rise to
a family of exact soliton solutions---see Eq. (\ref{peakon}) below---which
are completely unstable, but may be stabilized if the point-like
nonlinearity is combined with a periodic linear potential (in the same
setting, the localized nonlinearity with the repulsive sign may support
stable gap solitons) \cite{Nir}. In the present work, we demonstrate that
nonlinear self-trapped states may also be stabilized, without adding any linear
potential, by the regularized localized nonlinearity, if the ideal $\delta $%
-function is replaced by a Gaussian profile.

In addition to supporting stationary trapped modes, the localized potentials
may be used as \textit{tweezers}, for the transfer of the trapped modes,
which is a topic with many potential applications \cite{we}. In particular,
the use of the $\delta $-functional linear trapping potentials makes it
possible to find exact solutions for various transfer problems \cite%
{Edik,Adolfo}. The dynamical process of extracting matter-wave pulses from a
BEC reservoir by nonlinear tweezers, induced by an appropriate laser beam,
was simulated in detail in Ref.~\cite{Humberto}.

The main objective of the present work is to analyze possibilities of the
controlled transfer of localized wave packets by a moving nonlinear trapping
pseudopotential (\textit{nonlinear tweezers}), represented by a narrow
Gaussian profile multiplying the self-focusing cubic nonlinearity. An
advantage of using the nonlinear tweezers is that they would not grab and
drag small-amplitude (radiation)
``garbage", surrounding the target object in a
realistic setting, and one may use the amplitude of the object to control
its transfer (this resembles the well-known advantage of nonlinear optical
amplifiers in comparison with their linear counterparts---see Ref.~\cite%
{nonlin} and references therein). The model is introduced in Section 2,
where we also produce simple analytical results for the transfer problem,
obtained by means of the adiabatic approximation, which is appropriate for a
slowly moving trap. Basic numerical results are reported in Section 3. We
also briefly consider a related problem of dragging wave patterns with a
divergent norm by the nonlinear trap moving at a constant acceleration,
which is relevant in connection to the recently considered transmission
regimes for Airy beams, exact \cite{Berry} or approximate \cite{Airy},
including their nonlinear generalizations \cite{Airy-nonlinear}, and the
transfer of linear trapped wave packets by a potential well moving at a
constant acceleration \cite{Er'el}. In Section 4, we report results for
another natural problem related to the present setting, namely, the
scattering of linear wave packets on the localized stationary attractive
nonlinearity. Section 5 concludes the paper.

\section{The model and adiabatic approximation}

\subsection{The formulation}

The model with the tightly localized nonlinear trap, which represents the
moving tweezers, is based on the following normalized version of the NLS
equation for wave function $\psi (x,t)$:
\begin{equation}
i\psi _{t}=-\frac{1}{2}\psi _{xx}-\tilde{\delta}\left( x-\xi (t)\right)
\left\vert \psi \right\vert ^{2}\psi ,  \label{f}
\end{equation}
where subscripts denote partial derivatives, while
$\tilde{\delta}(x)$ and $\xi (t)$ define the shape of the nonlinear
trap and its law of motion. In the
context of BECs, Eq.~(\ref{f}) is a
normalized version of the Gross-Pitaevskii equation, with time $t$ and
coordinate $x$ [note that in an early work~\cite{Azbel}, Eq. (\ref{f}) was introduced
as a model for tunneling of attractively interacting particles through a
junction]. In the optical model, $t$ is replaced by the propagation distance
($z$), while $x$ is the transverse coordinate in the respective planar
waveguide.

In the first version of the model, the nonlinear trap was taken in the form
of the ideal delta-function, $\tilde{\delta}\left( x\right) =\delta (x)$
\cite{Azbel}. In the present work, numerical results are reported for its
practically relevant Gaussian regularization, namely
\begin{equation}
\tilde{\delta}(x)=\left( 2\sqrt{\pi \epsilon }\right) ^{-1}\exp \left(
-x^{2}/\left( 4\epsilon \right) \right) ,  \label{regularized}
\end{equation}%
with a sufficiently small width $\epsilon $. The law of motion for the
moving trap will be taken as
\begin{equation}
\xi (t)=\left\{
\begin{array}{cc}
0, & \mathrm{at~}~t\leq 0, \\
\frac{1}{2}\Xi \left[ 1+\sin \left( \pi \frac{t-T/2}{T}\right) \right] , &
\mathrm{at~}~0<t<T, \\
\Xi & \mathrm{at~}~t\geq T,%
\end{array}%
\right.  \label{motion}
\end{equation}%
which implies that the trap is smoothly transferred from the initial
position, $\xi =0$, starting at $t=0$, to the final position, $\xi =\Xi $,
at $t=T$. Both the initial and final velocities corresponding to this law of
motion are zero, i.e., $d\xi /dt(t=0)=d\xi /dt(t=T)=0$.

\subsection{Pulse solutions}

The model with the quiescent trap ($\xi =0$), described by the ideal $\delta
$-function, allows one to construct solutions to Eq.~(\ref{f}) as a
combination of two solutions of the linear Schr\"{o}dinger equation in free
space at $x<0$ and $x>0$, coupled by the derivative-jump condition at $x=0$
(while the wave function itself must be continuous at this point):
\begin{equation}
\psi _{x}(x=+0)-\psi _{x}\left( x=-0\right) =-2\left\vert \psi
(x=0)\right\vert ^{2}\psi (x=0).  \label{j}
\end{equation}%
This condition gives rise to a family of obvious solutions to Eq.~(\ref{f})
with $\tilde{\delta}(x)=\delta (x)$ and $\xi =0$:%
\begin{equation}
\psi =\left( -2\mu \right) ^{1/4}e^{-i\mu t}e^{-\sqrt{-2\mu }|x|},
\label{peakon}
\end{equation}%
where $\mu <0$ is an arbitrary chemical potential. Such soliton-like
solutions, with a discontinuous first derivative, are usually called \textit{%
peakons} (see, e.g., Refs.~\cite{camholm}). Note that the norm of the peakon
family \emph{does not} depend on the chemical potential, namely,
\begin{equation}
N\equiv \int_{-\infty }^{+\infty }\left\vert \psi (x)\right\vert ^{2}dx=1.
\label{1}
\end{equation}%
This degeneracy resembles the well-known feature of the \textit{Townes
solitons} in the two-dimensional NLS equation with the uniform nonlinearity
\cite{Berge}. Accordingly, the application of the Vakhitov-Kolokolov (VK)
stability criterion, $dN/d\mu <0$ \cite{VK,Berge}, formally predicts the
neutral stability of the peakon family; nevertheless, a numerical study
demonstrates that the entire family is unstable \cite{Nir}, the instability
being similar to that of the Townes solitons in two dimensions \cite{Berge}.
As mentioned above, the peakons can be stabilized by the addition of the
linear periodic potential (with the nonlinear $\delta $-function placed at
an arbitrary position with respect to the potential \cite{Nir}). In the next
section, we demonstrate that stabilization may also be induced by the
regularization of the $\delta $-function, as per Eq.~(\ref{regularized}).
Generally speaking, it is easy to stabilize the family of Townes-like
solitons, because the linearization of the underlying equation with respect
to small perturbations around the soliton does not give rise to any unstable
eigenvalue; in fact, the instability is nonlinear, i.e., it grows not
exponentially, but rather as a power of time, and is represented by a
respective pair of zero eigenvalues. Therefore, any modification of the
equation which shifts the zero eigenvalues in the direction of real
frequencies may stabilize the entire family. This will be the case in our
considerations below (in Section 3) with the finite-width Gaussian (\ref%
{regularized}) replacing the $\delta $-function.

Assuming that a stabilization mechanism is in action, but the shape of the
soliton does not substantially deviate from Eq.~(\ref{peakon}), it is
natural to expect that the trap, slowly moving according to given $\xi (t)$
[see, e.g., Eq.~(\ref{motion})], may \emph{drag} the trapped soliton, which
will then be described by the following modification of solution (\ref%
{peakon}), in the adiabatic approximation:
\begin{equation}
\psi =\left( -2\mu \right) ^{1/4}\exp \left[ -i\mu t+\frac{i}{2}\int \left(
\frac{d\xi }{dt}\right) ^{2}dt+i\frac{d\xi }{dt}x\right] \exp \left[ -\sqrt{%
-2\mu }\left\vert x-\xi (t)\right\vert \right] .  \label{adia}
\end{equation}%
In fact, expression (\ref{adia}) is the usually defined \textit{Galilean
boost} of peakon (\ref{peakon}) moving at instantaneous velocity $d\xi /dt$.

\subsection{Airy waves}

For the consideration of the nonlinear trap moving with constant
acceleration $W$, we set $\xi (t)=\left( W/2\right) t^{2}.$ Then, Eq. (\ref%
{f}) can be rewritten in the accelerating reference frame moving along with
the trap, i.e., in terms of the following variables:
\begin{eqnarray}
\psi \left( x,t\right) &\equiv &\left( W/4\right) ^{1/6}\phi \left( z,\tau
\right) \exp \left[ iWt\left( x-Wt^{2}/3\right) \right] , \\
z &\equiv &\left( 2W\right) ^{1/3}\left( x-Wt^{2}/2\right) ,\quad \tau
\equiv \left( W^{2}/2\right) ^{1/3}t.
\end{eqnarray}%
The accordingly transformed Eq.~(\ref{f}) reads:
\begin{equation}
i\phi _{\tau }+\phi _{zz}-z\phi +\delta \left( z-z_{0}\right) \left\vert
\phi \right\vert ^{2}\phi =0,  \label{phi}
\end{equation}%
where $z_{0}$ is the location of the nonlinear trap with respect to the
accelerating reference frame. Stationary solutions to Eq.~(\ref{phi}) are $%
\phi (z,\tau )\equiv a(z)$, with real function $a(z)$ obeying the following
equation:
\begin{equation}
\frac{d^{2}a}{dz^{2}}-za+\delta \left( z-z_{0}\right) a^{3}=0.  \label{Airy}
\end{equation}

Without the nonlinear term, Eq.~(\ref{Airy}) is the classical Airy equation,
whose relevant solutions at $z<z_{0}$ and $z>z_{0}$ are, respectively, given
by:
\begin{equation}
a_{-}(z)=C_{1}\mathrm{Ai}(z)+C_{2}\mathrm{Bi}(z),~a_{+}(z)=C_{3}\mathrm{Ai}%
(z).  \label{-}
\end{equation}%
Here, $C_{1}$ and $C_{2}$ are constants, and $\mathrm{Ai}(z)$, $\mathrm{Bi}%
(z)$ are the standard Airy functions which are defined by their asymptotic
forms at $z\rightarrow -\infty $ and $z\rightarrow +\infty $ \cite{Abram}:
\begin{equation}
\left\{ \mathrm{Ai,Bi}\right\} (z)\approx \frac{1}{\sqrt{\pi }\left(
-z\right) ^{1/4}}\left\{ \sin ,\cos \right\} \left( \frac{\pi }{4}+\frac{2}{3%
}\left( -z\right) ^{3/2}\right) ,  \label{Bi}
\end{equation}%
\begin{equation}
\left\{ \mathrm{Ai,Bi}\right\} (z)\approx \frac{1}{2\sqrt{\pi }z^{1/4}}\exp
\left( \mp \frac{2}{3}z^{3/2}\right) ~.  \label{AiBi}
\end{equation}%
Using the Wronskian of the Airy functions,
which is equal to $1/\pi $ \cite{Abram}, the continuity condition for $\phi
(z)$ and jump relation (\ref{j}) at $z=z_{0}$ give coefficients $C_{1,2,3}$
[see Eq. (\ref{-})] in terms of the amplitude of the solution, $A\equiv
a(z=z_{0})$, and $z_{0}$ itself, that may be considered as two free
parameters of the solution family:
\begin{eqnarray}
C_{1} &=&\left[ \left( \mathrm{Ai}(z_{0})\right) ^{-1}-\pi \mathrm{Bi}%
(z_{0})A^{2}\right] A,  \notag \\
C_{2} &=&\pi \mathrm{Ai}(z_{0})A^{3},  \notag \\
C_{3} &=&A\left( \mathrm{Ai}(z_{0})\right) ^{-1}.  \label{C2}
\end{eqnarray}

The known peculiarity of solutions based on the Airy functions is the
divergence of the norm in the tail at $z\rightarrow -\infty $, as the
corresponding average density produced by Eqs.~(\ref{-}) and (\ref{Bi}),
\begin{equation}
\left\langle a^{2}(z)\right\rangle \approx \frac{C_{1}^{2}+C_{2}^{2}}{2\pi
\sqrt{-z}},  \label{asympt}
\end{equation}%
decays too slowly. In time-dependent settings, the divergence of the norm
implies a gradual decay of the initial pulse, which
loses its power
sucked into the tail growing toward $z\rightarrow -\infty $ \cite{Airy}.

\section{Numerical results for the transfer and dragging problems}

\subsection{Stability conditions for the transfer}

First, we have constructed a family of stationary solutions to Eq.~(\ref{f})
with the quiescent ($\xi =0$) regularized nonlinearity profile, taken as per
Eq.~(\ref{regularized}), in the form of $\psi (x,t)=e^{-i\mu t}w(x)$, where
the real function $w(x)$ satisfies the equation:
\begin{equation}
\mu w=-\frac{1}{2}\frac{d^{2}w}{dx^{2}}-\tilde{\delta}(x)w^{3}.  \label{ODE}
\end{equation}

\begin{figure}[b]
\subfigure[]{\scalebox{0.35}{\includegraphics{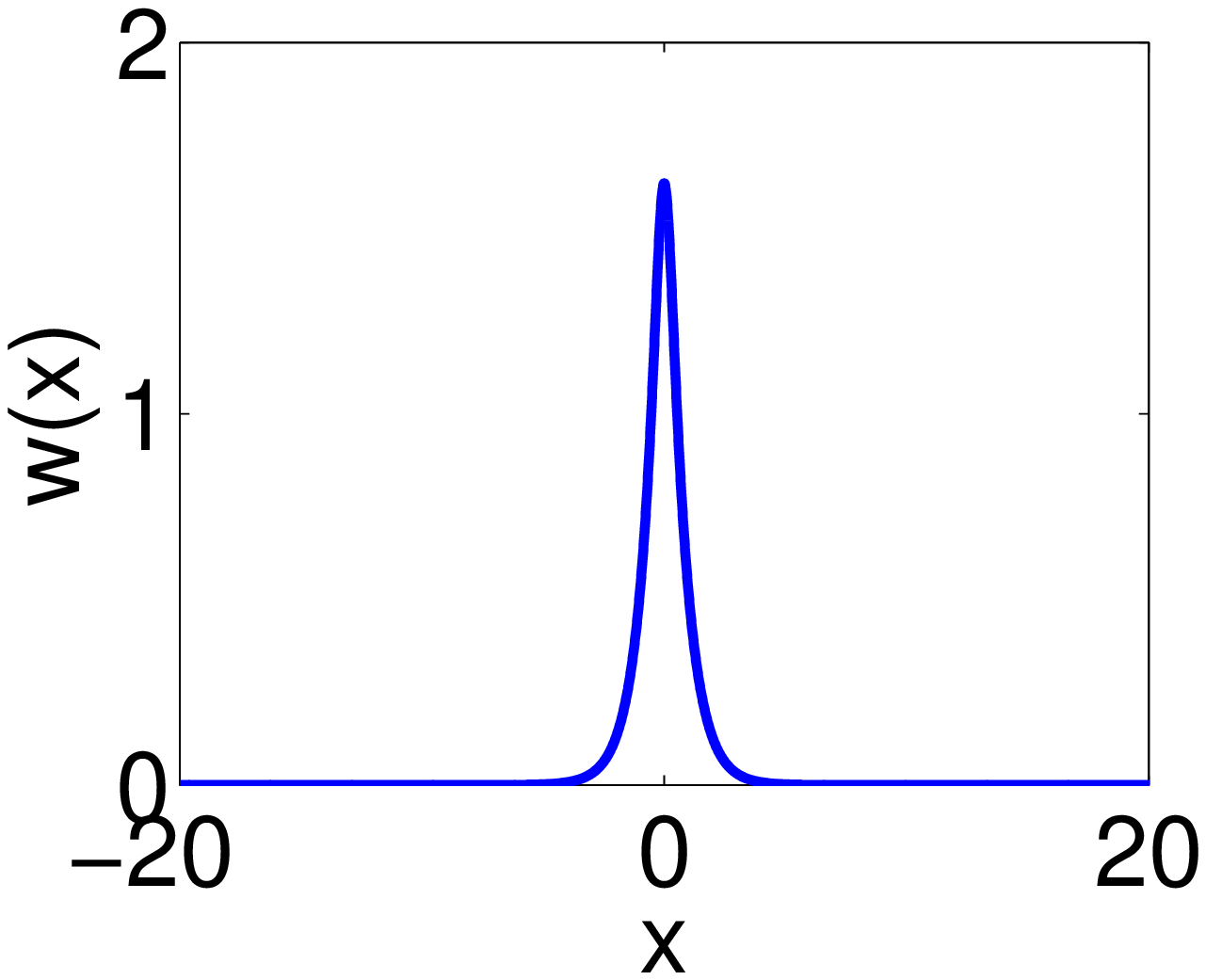}}} %
\subfigure[]{\scalebox{0.35}{\includegraphics{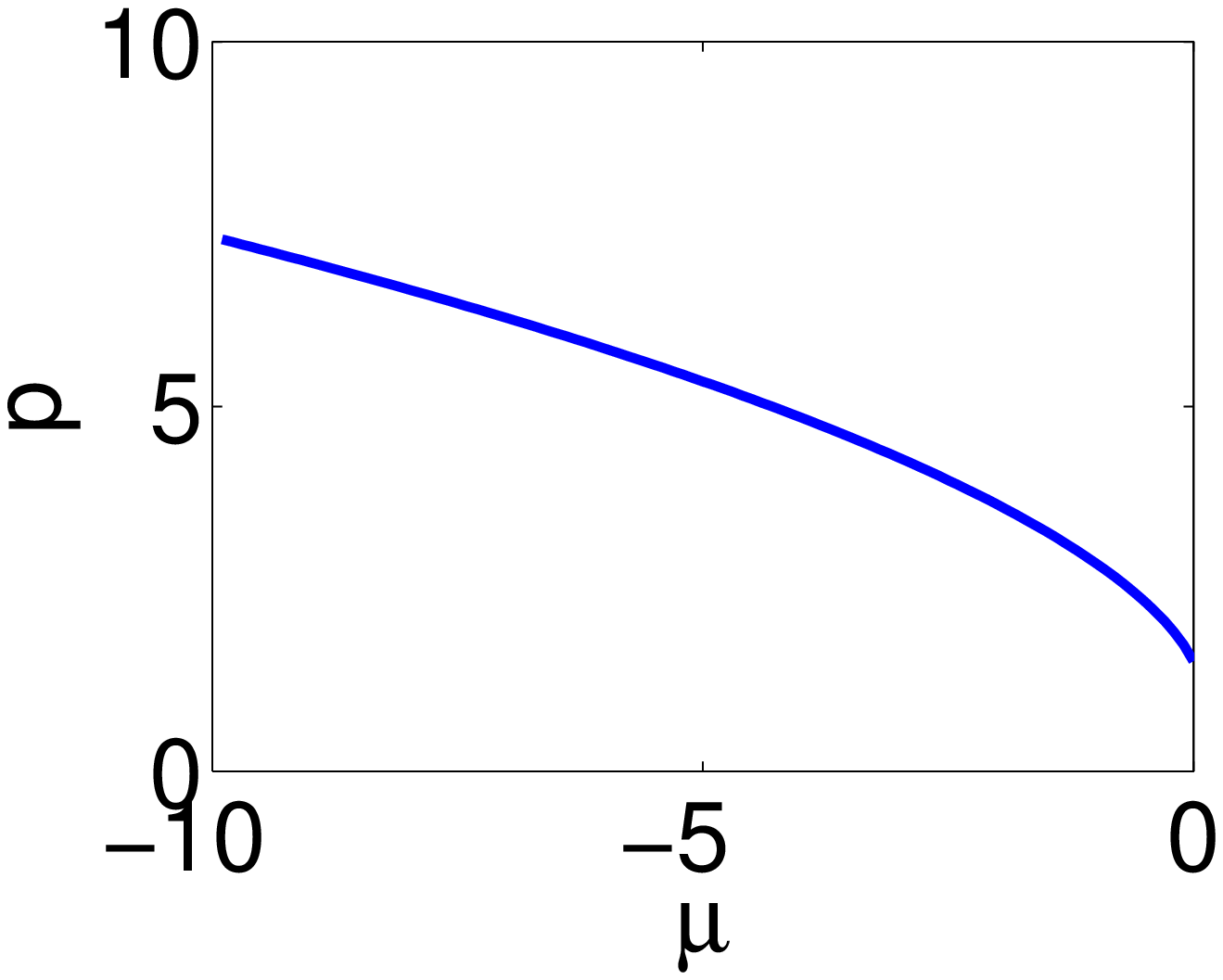}}} %
\subfigure[]{\scalebox{0.35}{\includegraphics{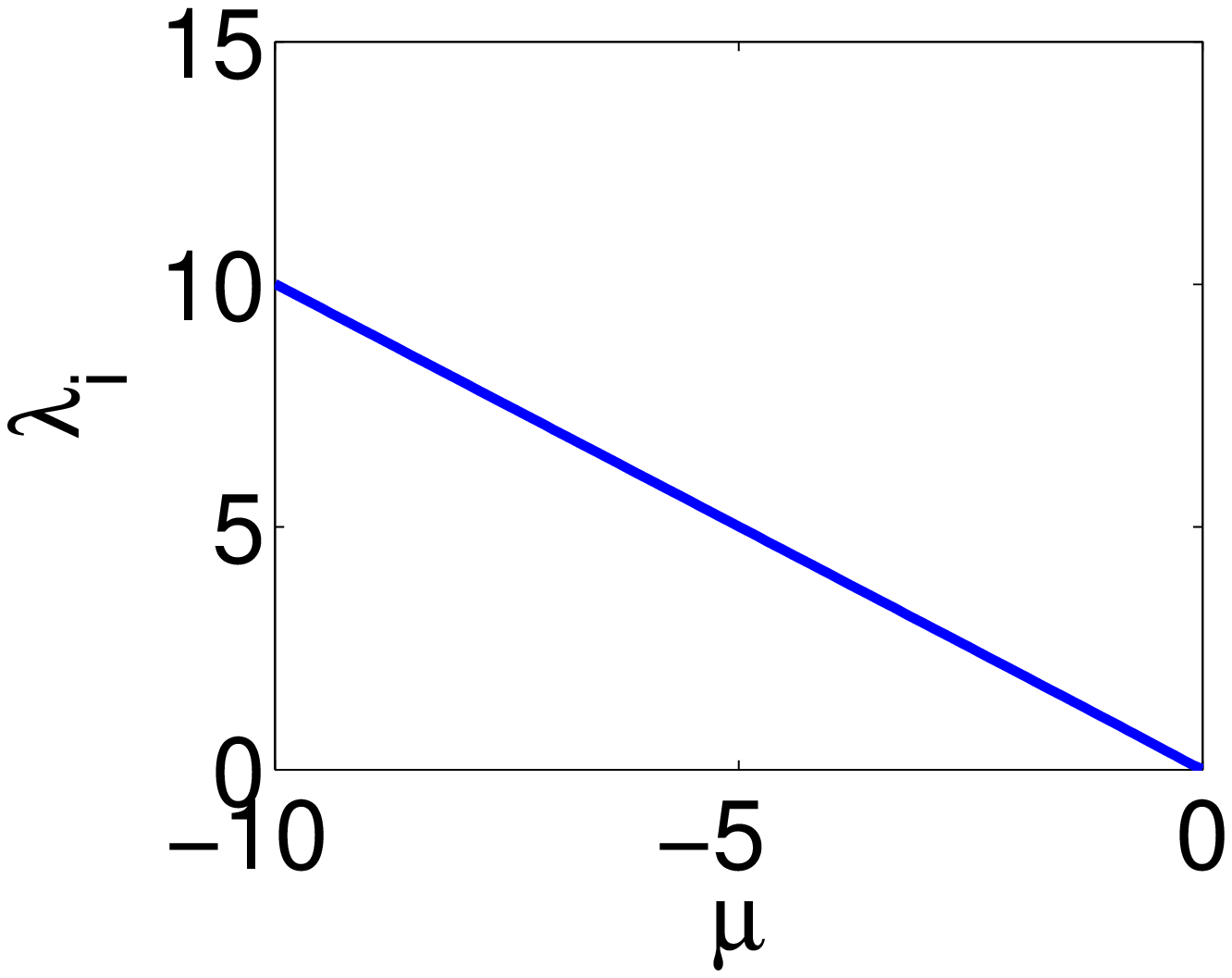}}}
\caption{(Color online) (a) An example of a numerically found stationary
solution of Eq.~(\protect\ref{ODE}) with $\protect\mu =-1$ and $\tilde{%
\protect\delta}(x)$ taken per Eq.~(\protect\ref{regularized}). (b) The norm
of the solutions, defined according to Eq.~(\protect\ref{1}), vs. chemical
potential $\protect\mu $. (c) The edge of the continuous spectrum lying on
the imaginary axis [the full eigenvalue is defined as $i\protect\lambda _{%
\mathrm{i}}$, the continuous spectrum occupying the area above the border
shown in (c)]. These results are obtained for a width of the moving trap of $%
\protect\epsilon =0.05$.}
\label{fig1}
\end{figure}

A typical example of the solution is displayed in Fig.~\ref{fig1}(a), and
the entire family of solutions is represented in Fig.~\ref{fig1}(b) by the
corresponding $N(\mu )$ curve, cf. the degenerate dependence given by Eq.~(%
\ref{1}) for the ideal $\delta $-function. It is obvious that the entire
family satisfies the VK criterion, $dN/d\mu <0$, hence the solitons may be
stable. This conjecture has been verified through the computation of the
stability eigenvalues for perturbations around the steady-state solution of
Eq.~(\ref{ODE}), and also by means of direct simulations of the perturbed
evolution (not shown here). All the eigenvalues were found to be purely
imaginary (they are defined so that imaginary values correspond to the real
frequencies, i.e., neutral stability), therefore only the edge of the
(continuous) spectrum \footnote{%
The spectrum is purely continuous, aside from a pair of eigenvalues at the
origin, which exist due to the $\mathrm{U}(1)$ invariance of the model.}
lying on the imaginary axis is shown in Fig.~\ref{fig1}(c). This
stabilization is a direct result of the regularization of the $\delta $%
-function in Eq. (\ref{ODE}). We have checked that, as smaller values of the
regularization parameter, $\epsilon $, are taken, the corresponding
stationary wave function $w$ approaches the peakon profile of Eq.~(\ref%
{peakon}). It is worthy to note that, as the limit of the ideal $\delta $%
-function is approached, we observe a pair of eigenvalues that bifurcate
from the edge of the continuous spectrum and approach the origin, where they
formally arrive in the limit of $\epsilon \rightarrow 0$, at which point the
scaling invariance associated with Eq.~(\ref{1}) is restored, and the
nonlinear instability emerges.

\begin{figure}[b]
\subfigure[]{\scalebox{0.28}{\includegraphics{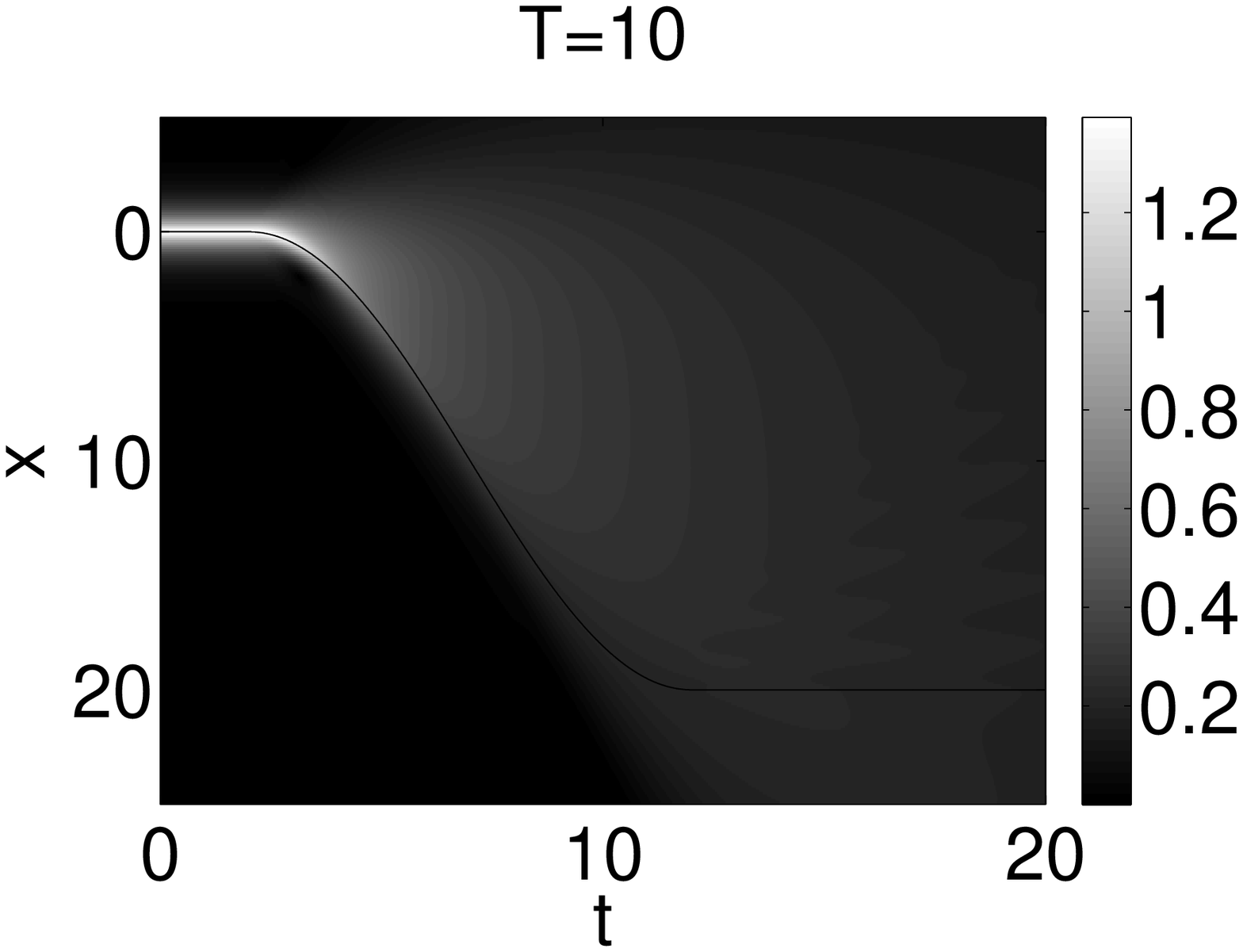}}}%
\subfigure[]{\scalebox{0.28}{\includegraphics{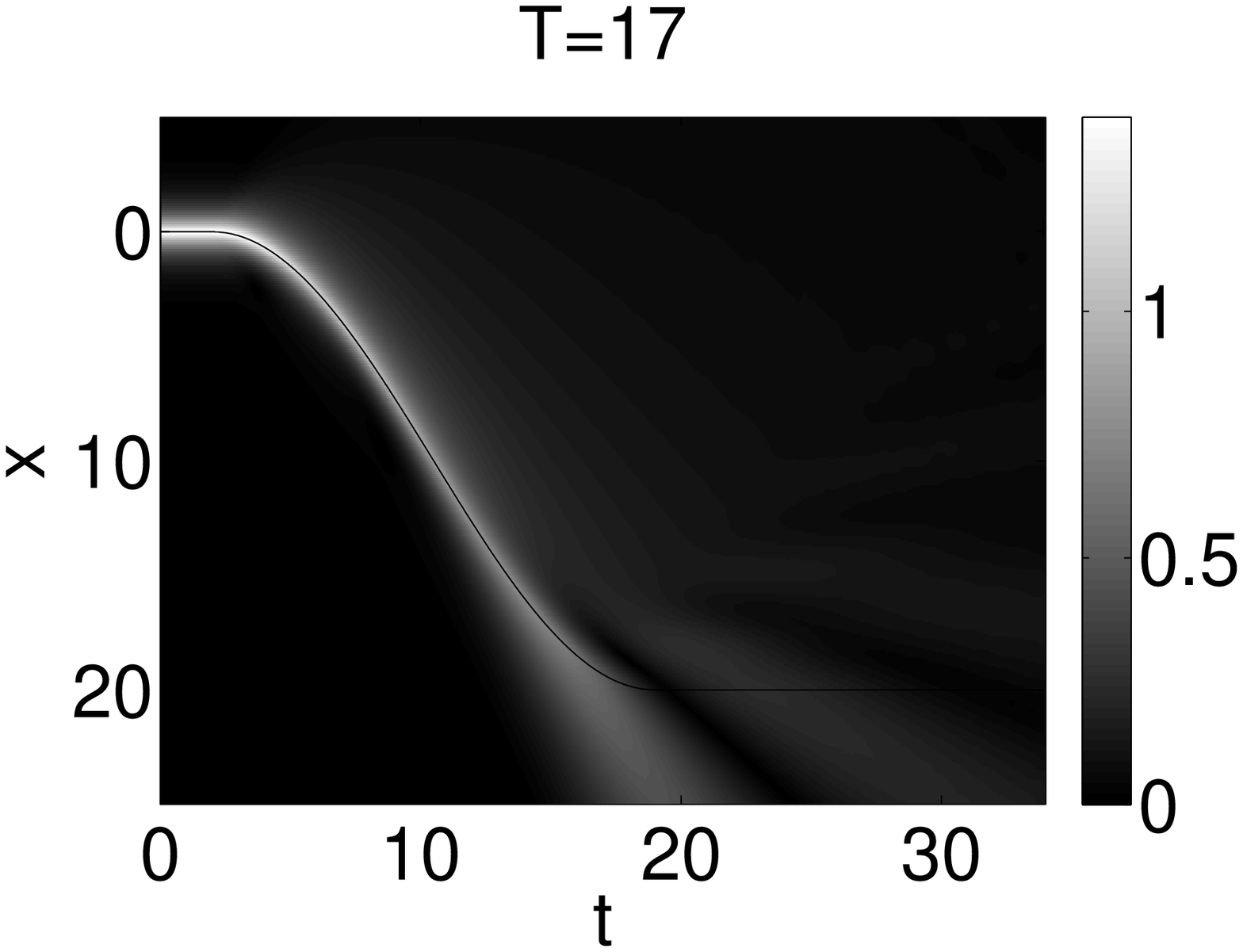}}}%
\subfigure[]{\scalebox{0.28}{\includegraphics{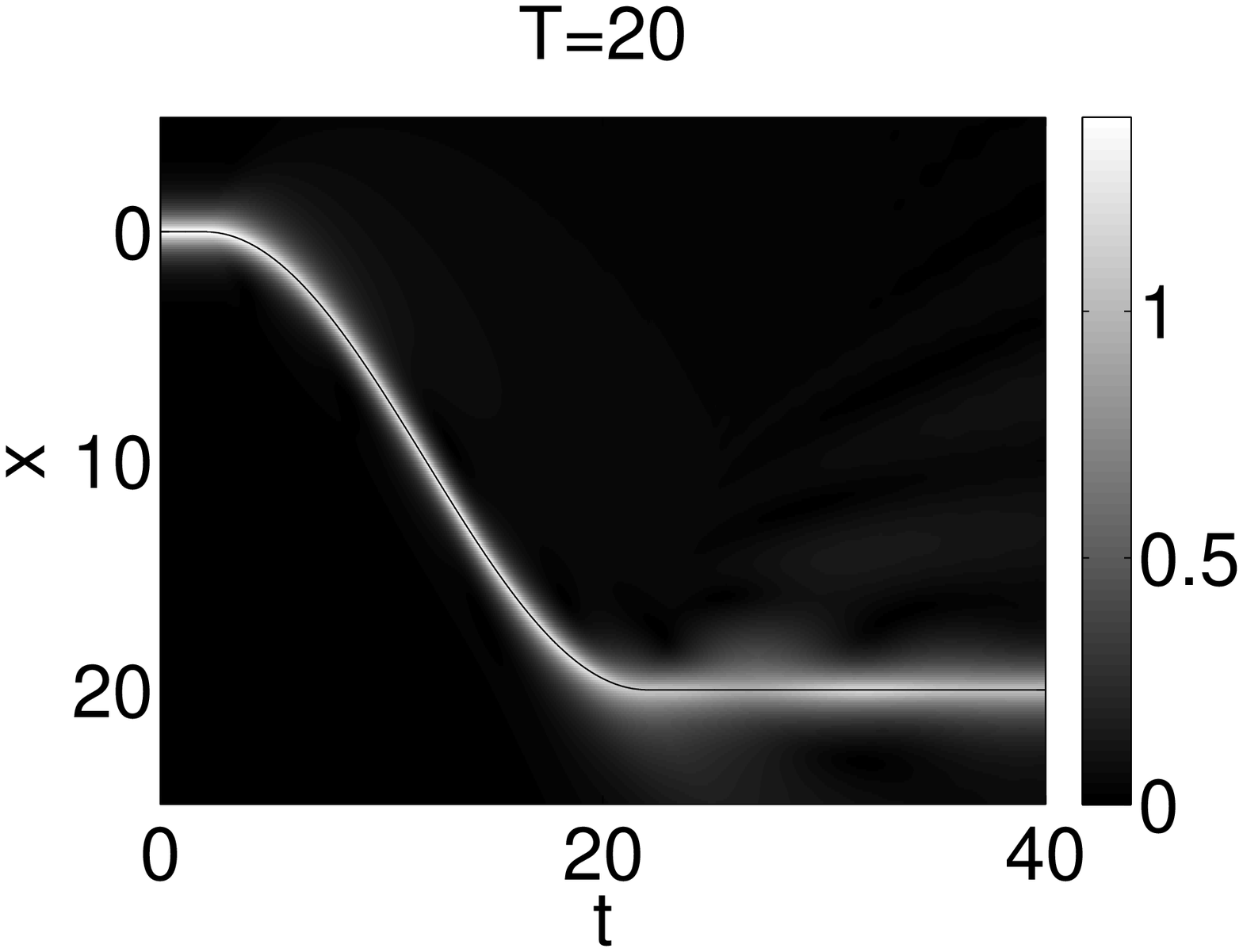}}} %
\subfigure[]{\scalebox{0.38}{\includegraphics{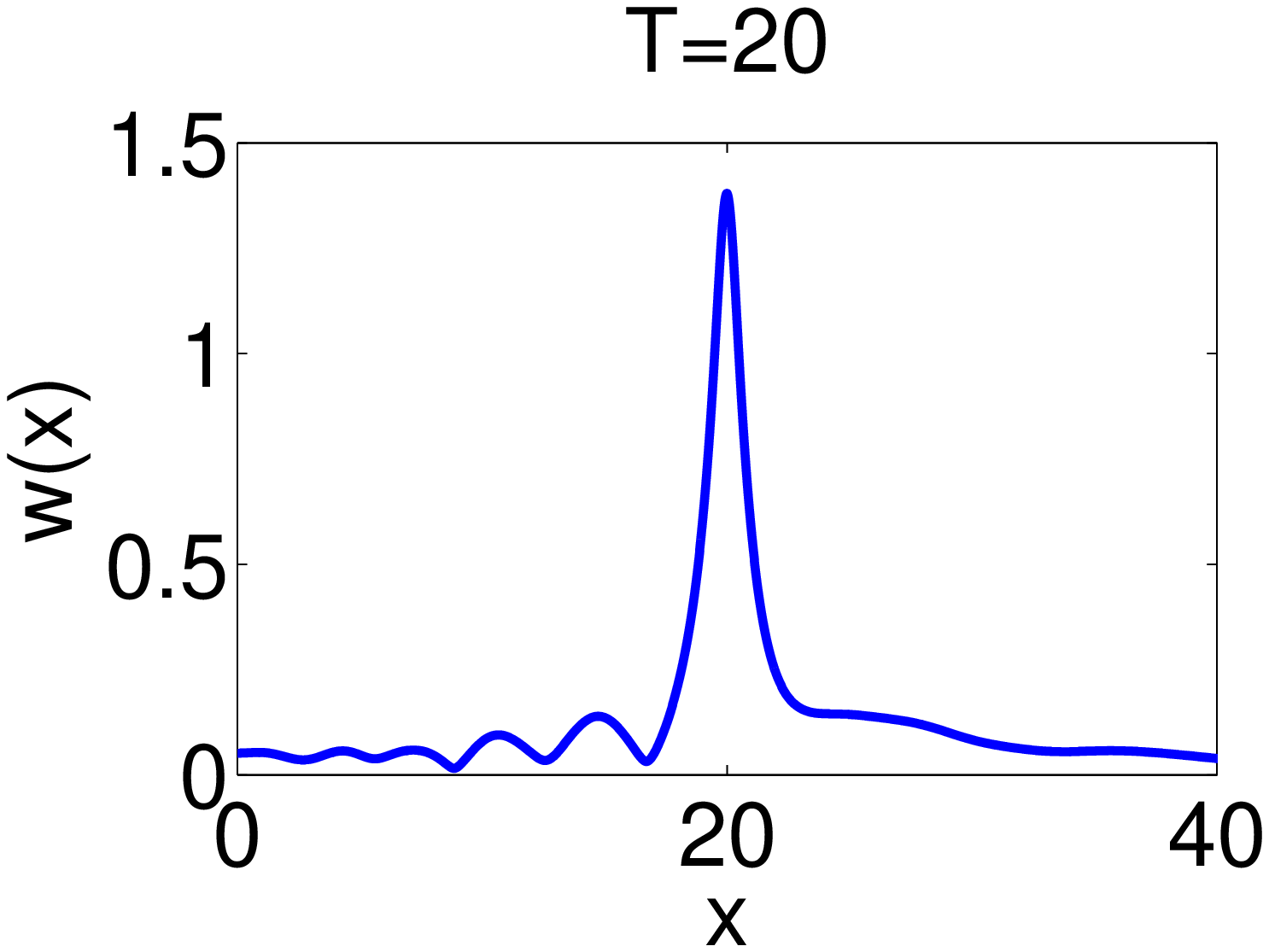}}}\subfigure[]{%
\scalebox{0.4}{\includegraphics{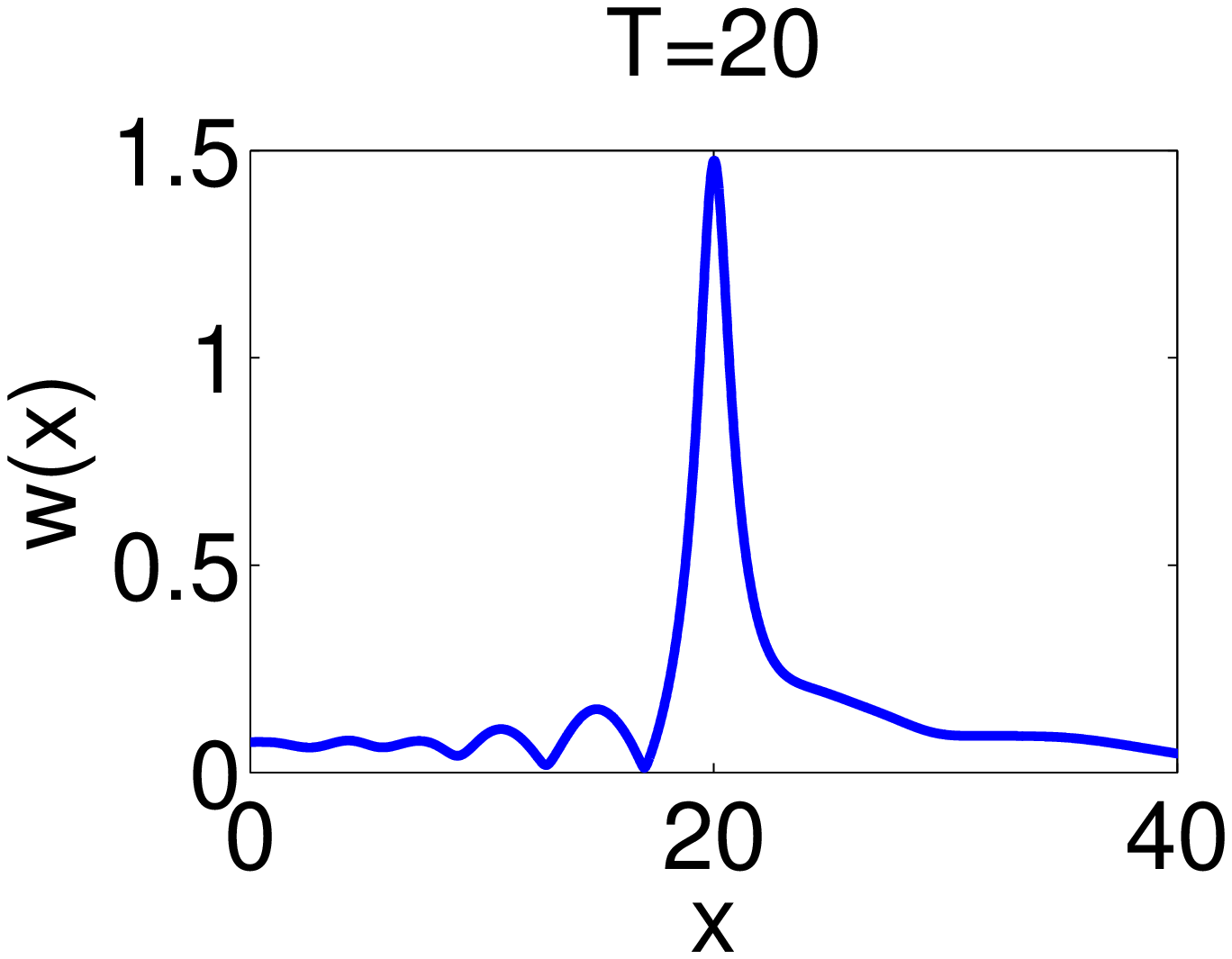}}}\subfigure[]{\scalebox{0.4}{%
\includegraphics{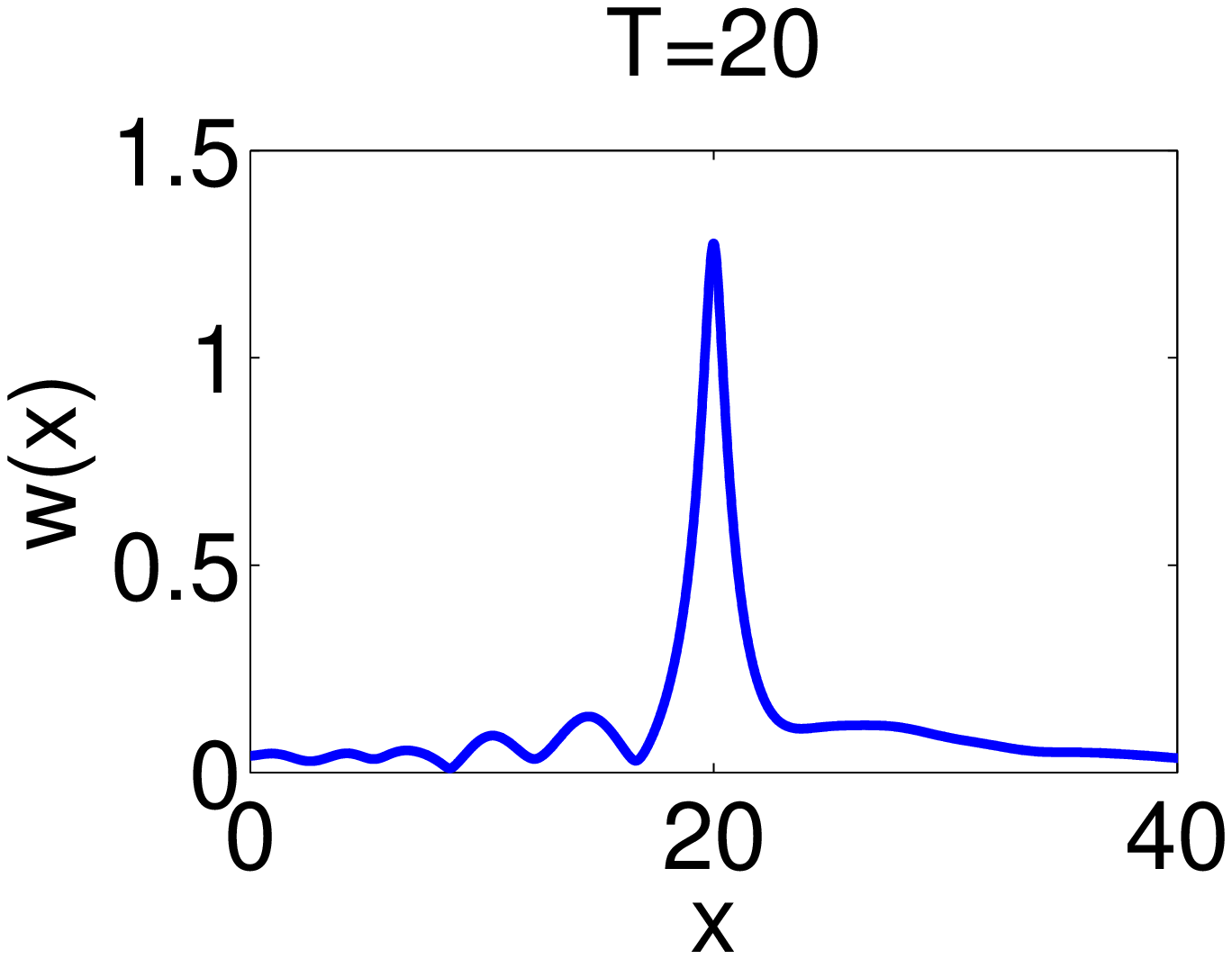}}}
\caption{(Color online) (a,b,c): Density contour plots illustrating the
dragging of the soliton, shown in Fig.~\protect\ref{fig1}(a), by the
nonlinear trap with width $\protect\epsilon =0.05$, moving as per Eq.~(%
\protect\ref{motion}), for three different values of the dragging time, $T$.
The black solid lines show the corresponding trajectory $\protect\xi (t)$.
Adopting the criterion that the surviving soliton must keep $\geq 90\%$ of
its initial amplitude, we conclude that it survives the transfer for $T\geq
20$. Panel (d) displays the shape of the soliton in the case of $T=20$,
i.e., at the border of the survival, at time $t=30$ (some time after the
completion of the transfer). Panels (e) and (f) display the same as (d), but
for the moving trap with widths $\protect\epsilon =0.1$ and $\protect%
\epsilon =0.025$, respectively.}
\label{fig2}
\end{figure}

We now turn to characteristic examples (cf. Fig.~\ref{fig2}) of the transfer of
originally stable quiescent solitons by the nonlinear trap moving according
to Eq.~(\ref{motion}). As might be expected, the soliton survives the
transfer, provided that it is not dragged too hard, i.e., the transfer time,
$T$, is not too short. We define the soliton as ``surviving the transfer"
if, at time $t=T$, its amplitude is no smaller than a sufficiently high
fraction of the initial soliton amplitude---say, $90\%$ or $80\%$. For
instance, in the case shown in Fig.~\ref{fig2}, the soliton dragged by the
trap of width $\epsilon =0.05$ keeps $90\%$ of the initial amplitude for $%
T\geq 20$. It can be checked that, in all cases of the ``survival", the
motion of the dragged soliton accurately follows the prediction of the
adiabatic approximation [Eq.~(\ref{adia})], as is clearly seen in the
example displayed in Fig. \ref{fig2}(c), where the actual trajectory of the
dragged mode closely follows the adiabatic trajectory $\xi (t)$. The
surviving solitons essentially preserve their original shape of the smoothed
peakon, cf. Figs. \ref{fig1}(a) and \ref{fig2}(d,e,f). The three latter
figures demonstrate that the eventual shape of the transferred mode is
nearly the same for three different values of the trap's width, $\epsilon
=0.05$, $0.1$, and $0.025$, i.e., the eventual results are robust,
withstanding the variation of essential parameters of the scheme within wide
limits.

\begin{figure}[tph]
\scalebox{0.3}{\includegraphics{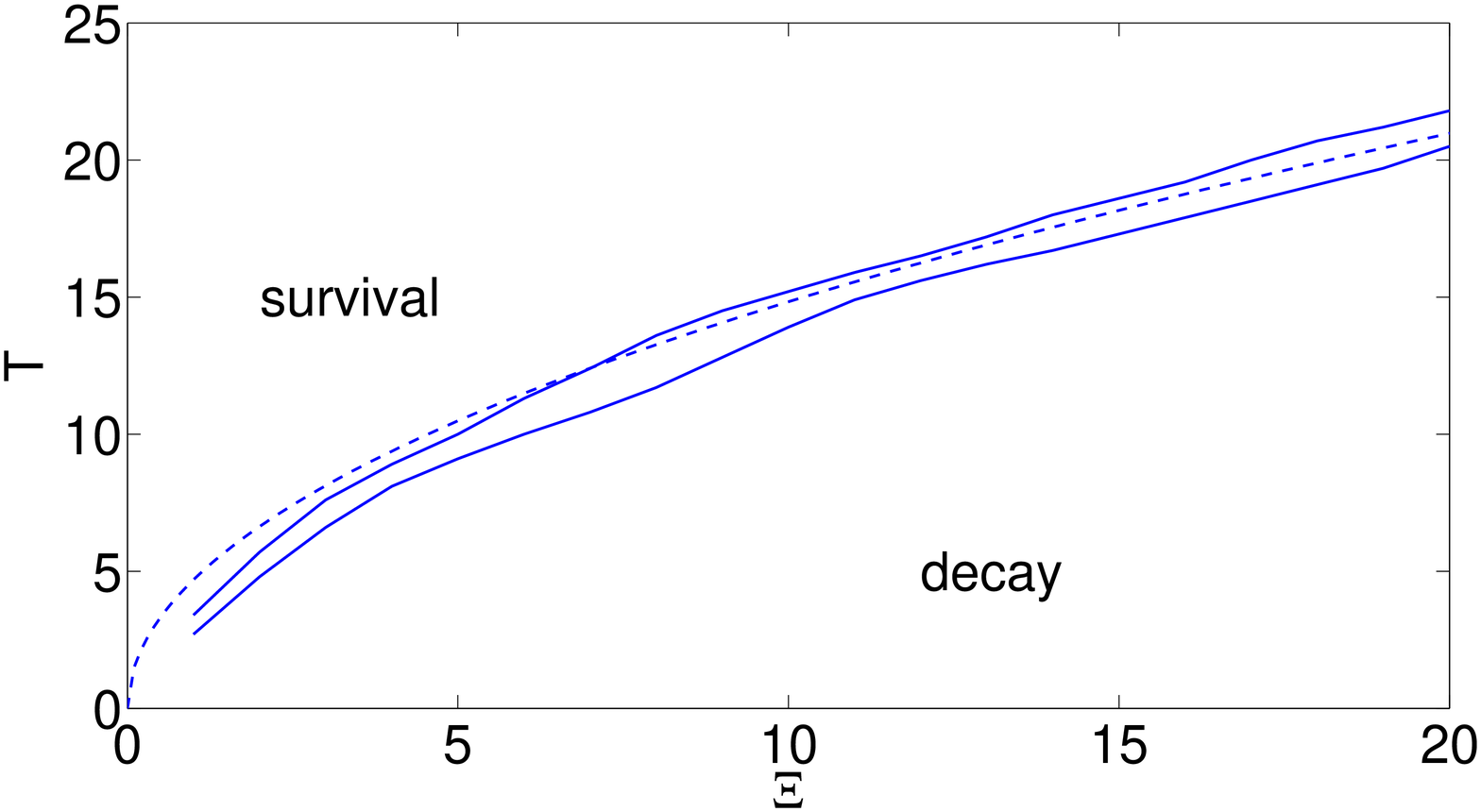}}
\caption{(Color online) The top and bottom solid curves represent the
boundary above which the initial soliton with chemical potential $\protect%
\mu =-1$, shown in Fig.~\protect\ref{fig1}(a), survives, keeping more than,
respectively, $90\%$ or $80\%$ of the initial amplitude by $t=30$. The
dashed curve depicts the analytical estimate (see text).}
\label{fig3}
\end{figure}

Results of the systematic simulations are summarized in Fig.~\ref{fig3},
where the survival border is shown in the plane of the transfer parameters, $%
\Xi $ and $T$ [see Eq. (\ref{motion})], for the fixed initial soliton and
two different survival criteria, based on the demand of keeping the
amplitude at the $90\%$ or $80\%$ level. It may again be concluded that the
results are robust, as they only weakly depend on the particular choice of
the criterion of the soliton's survival. The roughly parabolic shape of the
border may be explained by the fact that (as discussed in the previous
section) the gradual destruction of the dragged soliton is determined by the
absolute value of the acceleration of the moving trap. In the case of the
motion law (\ref{motion}), the acceleration is
\begin{equation}
\frac{d^{2}\xi }{dt^{2}}=-\frac{\pi ^{2}\Xi }{2T^{2}}\sin \left( \pi \frac{%
t-T/2}{T}\right) ,
\end{equation}%
hence its average absolute value in the process of the dragging is $%
\left\langle \left\vert d^{2}\xi /dt^{2}\right\vert \right\rangle =\pi \Xi
/T^{2}$. Assuming that the soliton survives when the absolute value of the
acceleration does not exceed a a certain critical level, the latter result
predicts a parabolic shape of the survival border, $T\sim \sqrt{\Xi }$. To
confirm the reasonable accuracy of the prediction, we have drawn a parabola
with an empirically fitted coefficient, $T=\sqrt{7\pi \Xi }$ (the dashed
line), to be compared to the numerically determined survival borders
displayed in Fig.~\ref{fig3}.

\subsection{Quasi-Airy profiles dragged at a constant acceleration}

\begin{figure}[tbp]
\subfigure[]{\scalebox{0.21}{\includegraphics{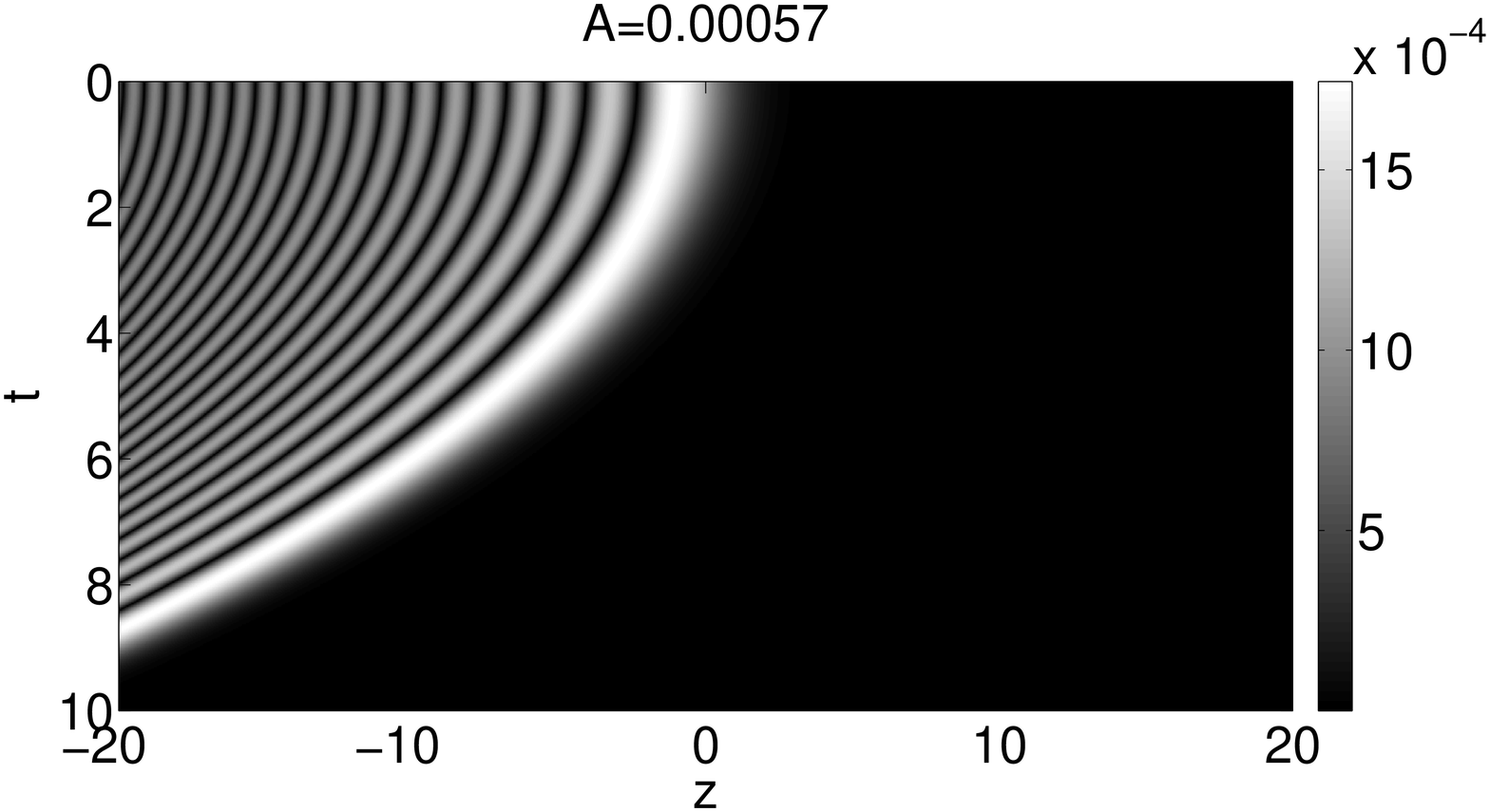}}} %
\subfigure[]{\scalebox{0.21}{\includegraphics{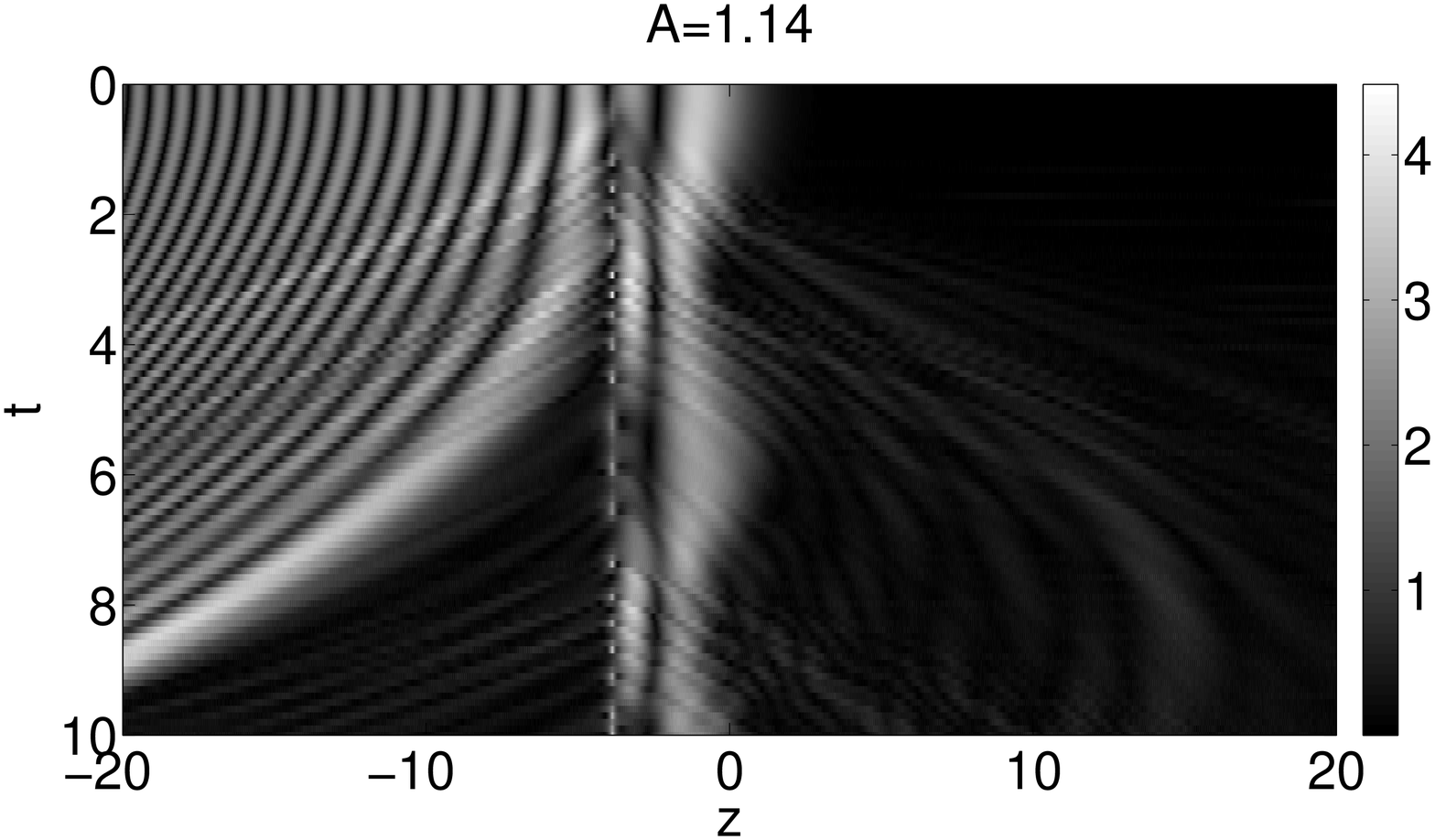}}}
\caption{The temporal evolution of $a(z)$ with $z_{0}=-3.86235$, for the
quasi-Airy waves dragged by the accelerating nonlinear trap with width $%
\protect\epsilon =0.05$. The left and right panels show, respectively, the
stationary solution with $A=0.00057$, and the unstable solution with $A=1.14$%
.}
\label{fig4}
\end{figure}

We have also briefly studied profiles of waves dragged at a constant
acceleration, i.e., those built as per Eqs. (\ref{-}) and (\ref{C2}), and
their stability within the framework of Eq. (\ref{phi}). A typical example
of the stationary wave attached to the accelerating nonlinear $\delta $%
-function, i.e., a solution to Eq. (\ref{Airy}), is displayed in Fig. \ref%
{fig4}. Direct simulations of Eq. (\ref{phi}) with this stationary wave used
as the initial configuration were performed in order to examine its
robustness. In particular, $a(z)$ was constructed and an approximate local
minimum of the quantity $C_{1}^{2}+C_{2}^{2}$ [which determines the tail's
amplitude, $\left\langle a^{2}(z)\right\rangle $, as per Eq.~(\ref{asympt})]
was identified for appropriate values of $A$ and $z_{0}$. The temporal
evolution of the resulting configuration is shown in Fig. \ref{fig4}.

A more detailed analysis of the quasi-Airy waves in the present nonlinear
model will be a subject of a separate work. We note here that if we use
small $A$ (such as $A=0.00057$ in the left column of Fig. \ref{fig4}), the
apparent discontinuity of the beam at $z_{0}$ is barely visible; in this
small-amplitude case, the solution appears to be more robust, and moves to
the left, preserving its shape. If $A$ is larger ($A=1.14$ in the right
panel), a significant fraction of the solution gets trapped by the
delta-function at $z_{0}$, while the larger remaining fraction still moves
to the left, and a smaller fraction of the initial power disperses to the
right of $z_{0}$.

\section{Scattering of coherent wave packets on the localized nonlinearity}

The scattering problem in the framework of Eq.~(\ref{f}) with the ideal
quiescent $\delta $-function was first studied in Ref.~\cite{Azbel}, where
the scattering of an incident plane wave was considered. It was found that
the localized self-attractive nonlinearity gave rise to an accordingly
localized modulational instability (MI)\ of the incident wave, provided that
its amplitude exceeded a certain threshold (minimum) value. Unlike the
commonly known MI in the NLS equation with the uniform nonlinearity, the
localized MI features a complex growth rate (i.e., it corresponds to the
oscillatory instability).

\begin{figure}[tph]
\subfigure[]{\scalebox{0.4}{\includegraphics{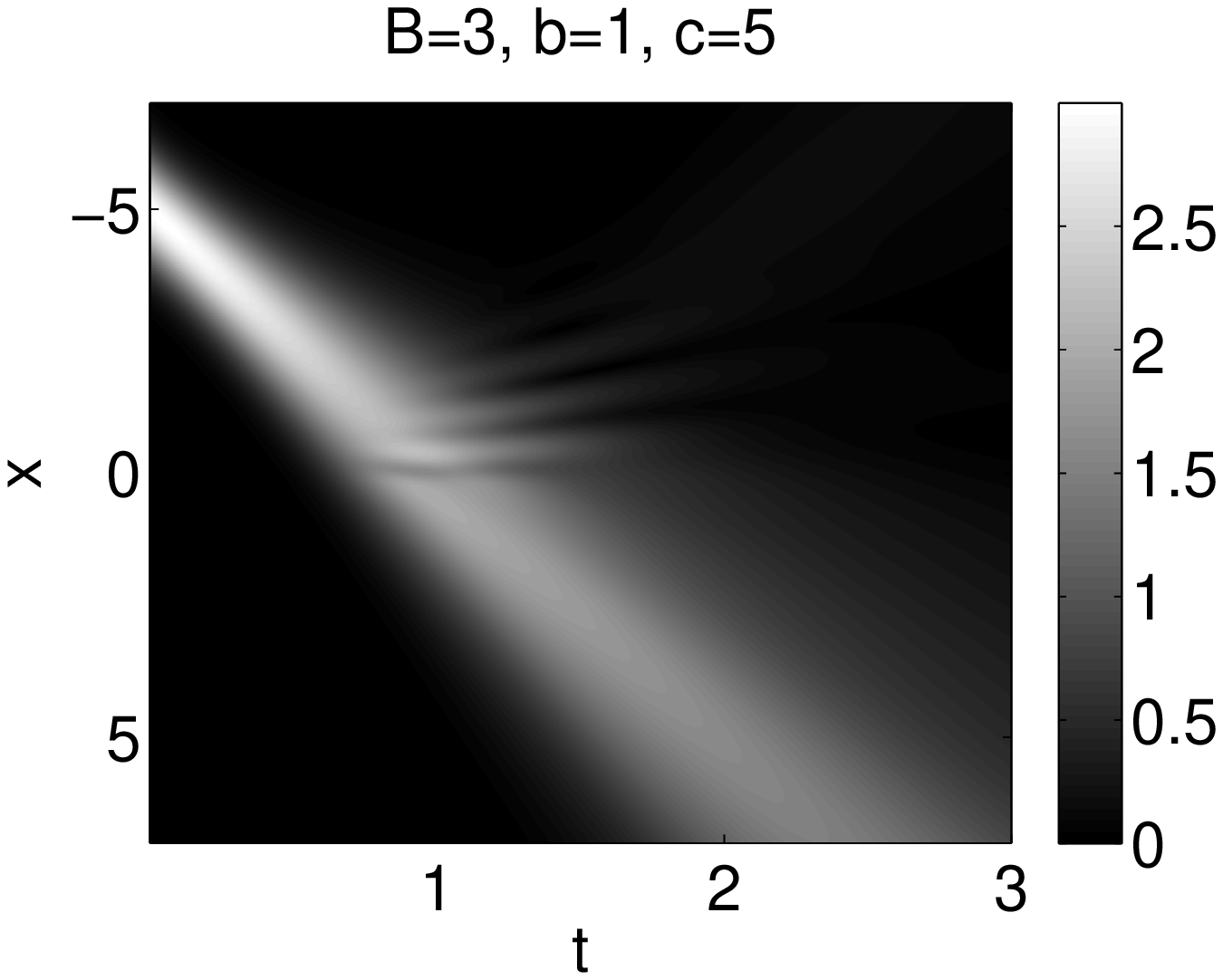}}} %
\subfigure[]{\scalebox{0.4}{\includegraphics{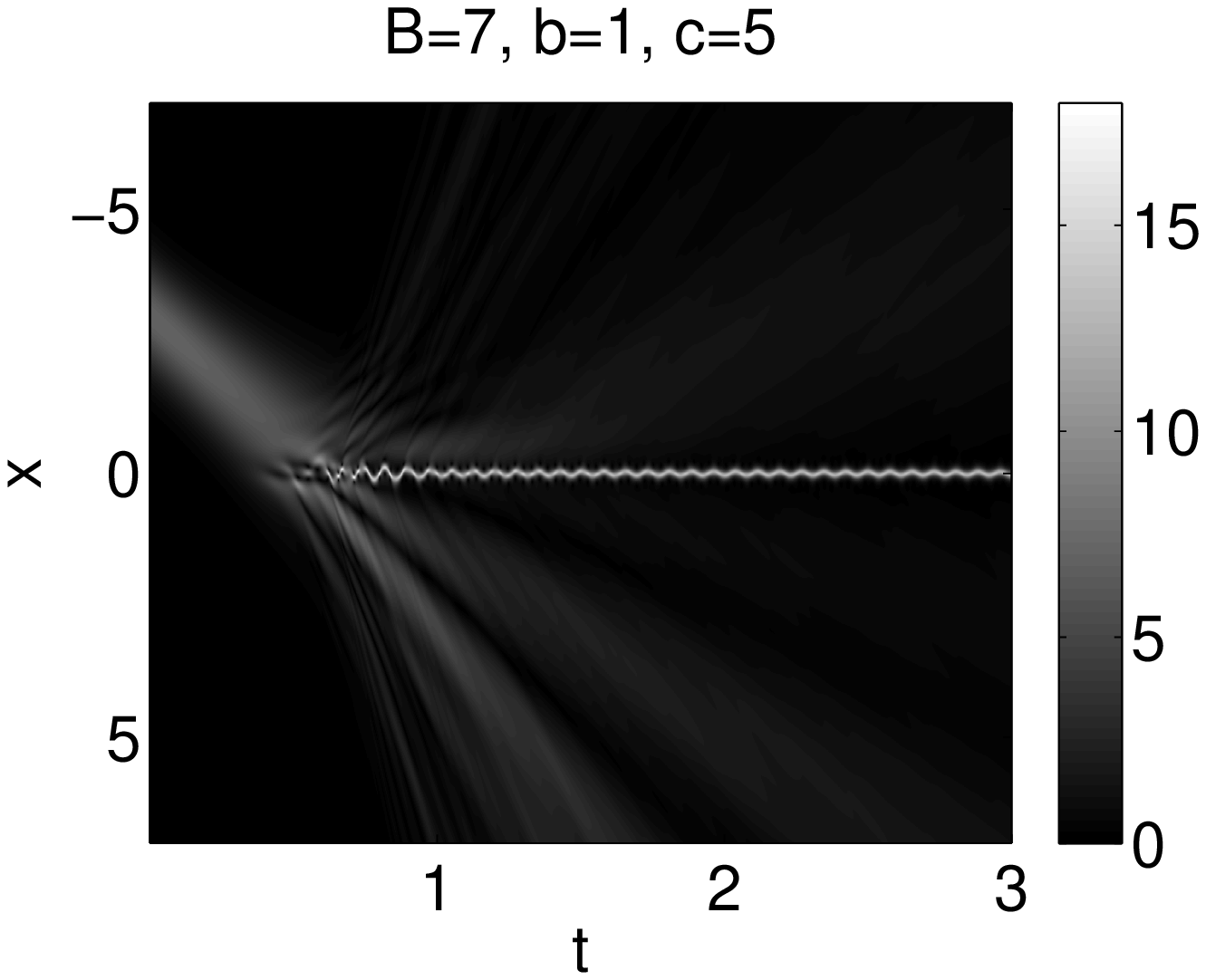}}} %
\subfigure[]{\scalebox{0.4}{\includegraphics{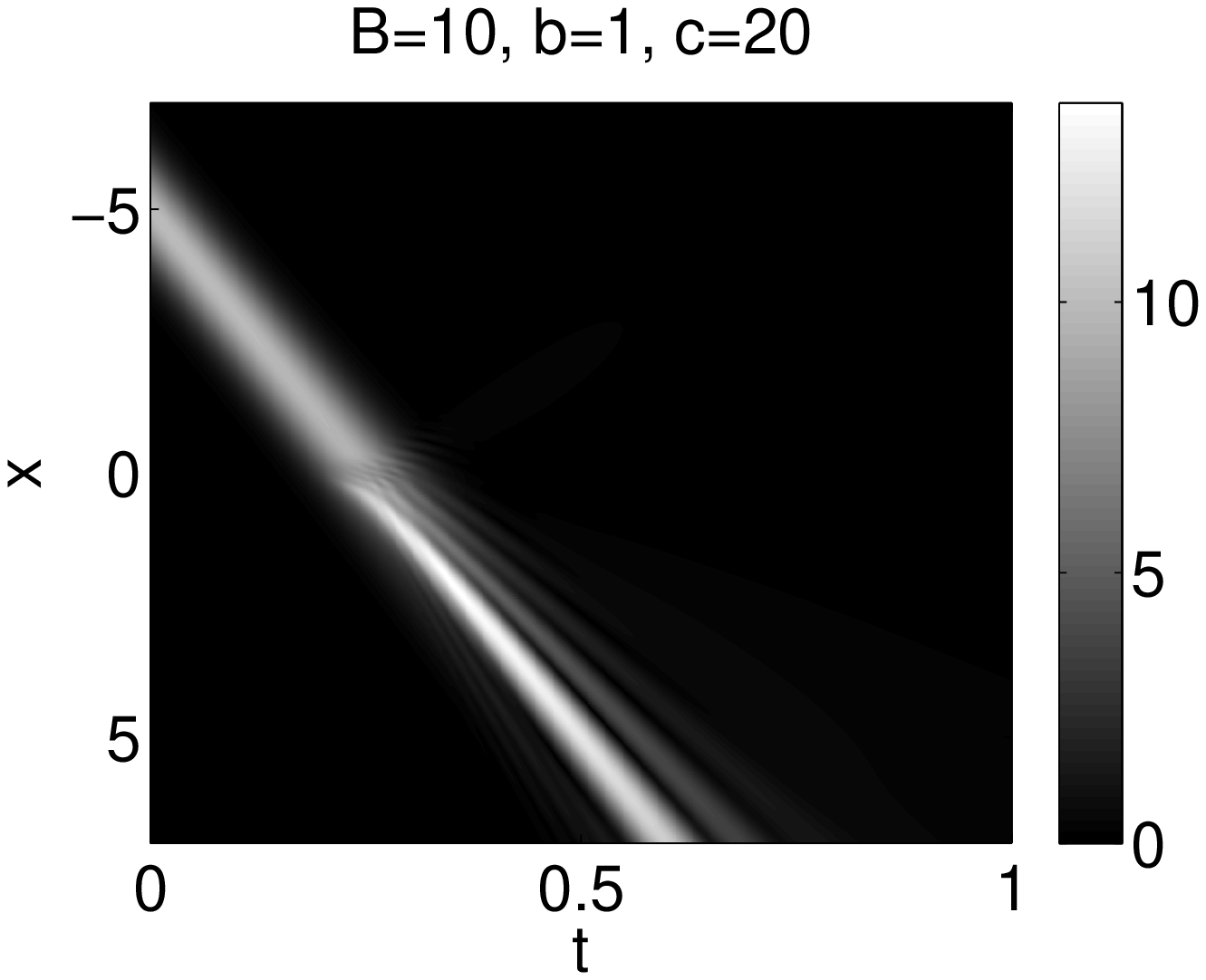}}} %
\subfigure[]{\scalebox{0.4}{\includegraphics{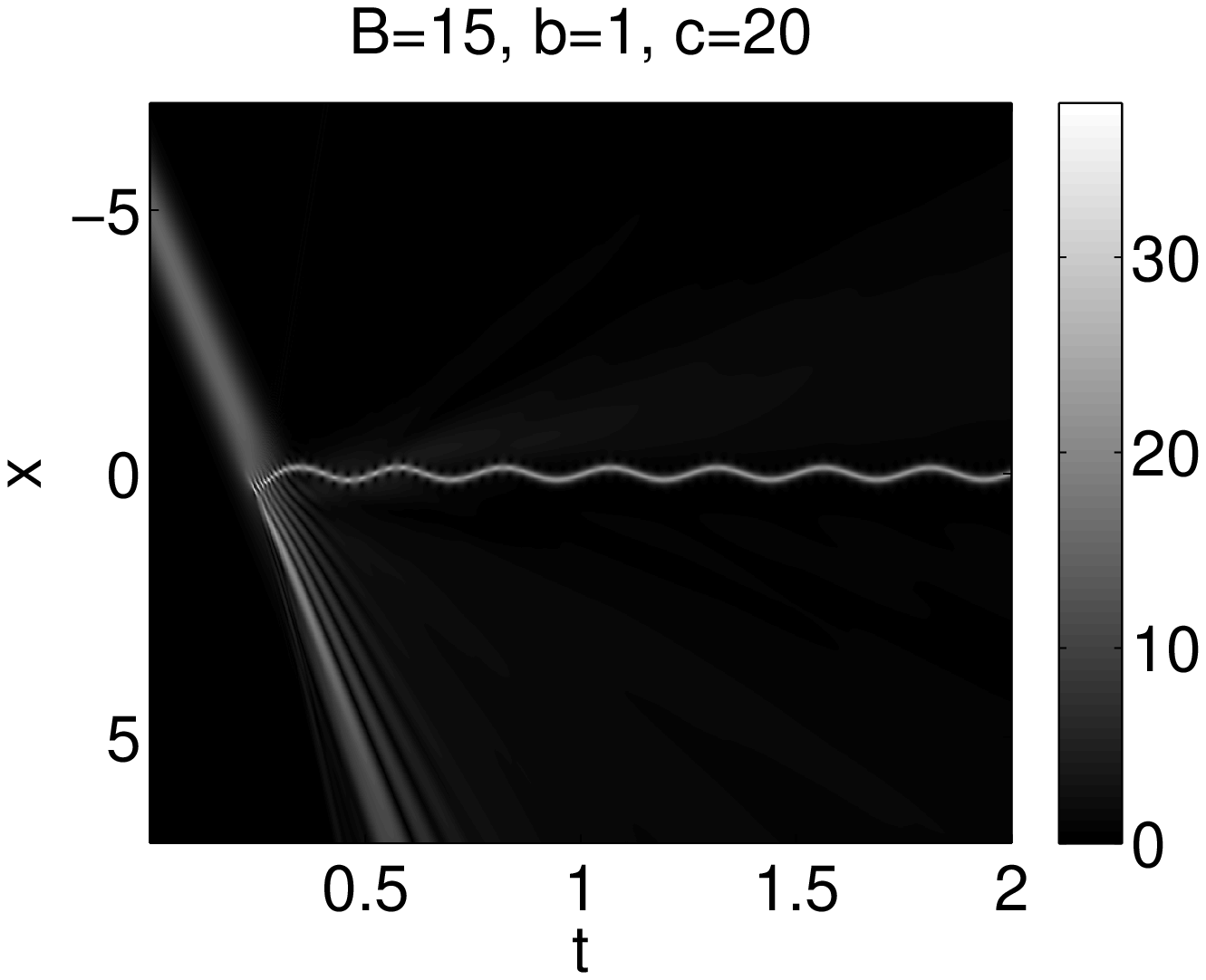}}}
\caption{Examples of the interaction of the incident coherent pulse (\protect
\ref{coh}) with the localized nonlinear potential corresponding to the
regularized profile (\protect\ref{regularized}), for $\protect\epsilon =0.05$%
, at different values of the pulse's amplitude and velocity.}
\label{fig5}
\end{figure}

Here, we aim to study the scattering from the regularized Gaussian nonlinear
potential of incident wave packets in the form of coherent states. The
latter represent the fundamental solutions of the linear Schr\"{o}dinger
equation in the free space,
\begin{equation}
\psi _{\mathrm{in}}(x,t)=\frac{B}{\sqrt{1+ibt}}\exp \left[ -\frac{b\left(
x-ct\right) ^{2}}{1+ibt}+icx+\frac{ic^{2}}{2}t\right] ,  \label{coh}
\end{equation}%
Here, $B$ and $b$ are real constants that determine, respectively, the
amplitude and inverse width of the initial wave packet, $A=\left(
1+b^{2}t^{2}\right) ^{-1/4}B$ and $W=\sqrt{b^{-2}+t^{2}}$ are the same
characteristics of the spreading packet; the norm of the packet is $N=\sqrt{%
\pi /(2b)}B^{2}$, while $c$ is its velocity. In particular, in the case of
small $B^{2}$, the scattering problem may be considered in the Born's
approximation (see, e.g., Ref.~\cite{saku}), by setting $\psi (x,t)=\psi _{%
\mathrm{in}}(x,t)+\psi _{\mathrm{Born}}\left( x,t\right) $, where the first
correction is to be found from the linear inhomogeneous equation,
\begin{equation}
i\left( \psi _{\mathrm{Born}}\right) _{t}+\frac{1}{2}\left( \psi _{\mathrm{%
Born}}\right) _{xx}=-\delta (x)\left\vert \psi _{\mathrm{in}}\right\vert
^{2}\psi _{\mathrm{in}}.  \label{Born}
\end{equation}%
Equation (\ref{Born}) can be solved by means of the Fourier transform. Here,
we instead focus  on direct simulations of the scattering problem in a
wide parametric region in the space of $(B,b)$, with the $\delta $-function
regularized according to Eq.~(\ref{regularized}). As illustrated by the
examples displayed in Fig.~\ref{fig5}, the simulations demonstrate splitting
of the incident wave packet into three parts---transmitted, trapped, and
reflected ones. Naturally, the share of the trapped norm increases with $B$,
i.e., with the enhancement of the nonlinearity; on the other hand, the
dependence of the observed phenomenology on the inverse-width parameter $b$
is relatively weak. It is also worthy to mention that, as seen in Fig.~\ref%
{fig5}(d), the trapped pulse performs oscillations around the localized
nonlinearity, while the transmitted beam splits into jets---see Figs.~\ref%
{fig5}(c,d).

Results of the simulations of the scattering are summarized in Figs. \ref%
{fig6} and \ref{fig7},
where we show the shares of the
trapped, reflected, and transmitted norm (or power, in terms of the optics
model), as functions of the amplitude ($B$) and inverse width ($b$) of the
incident pulse for a fixed velocity, $c=20$, and two different values of the
scatterer's width, $\epsilon =0.05$ and $0.1$, respectively. The plots
correspond to the moment of time $t=0.5$, taken after the the incident pulse
has passed the scatterer. Note that the sum of the three shares is, by
construction, identically equal to $1$. Analysis of the numerical data
demonstrates that, in the limit of large amplitude ($B$), the transmission
decreases while the reflection of the wave packet from the nonlinear barrier
gets enhanced.
\begin{figure}[tph]
\subfigure[]{\scalebox{0.38}{\includegraphics{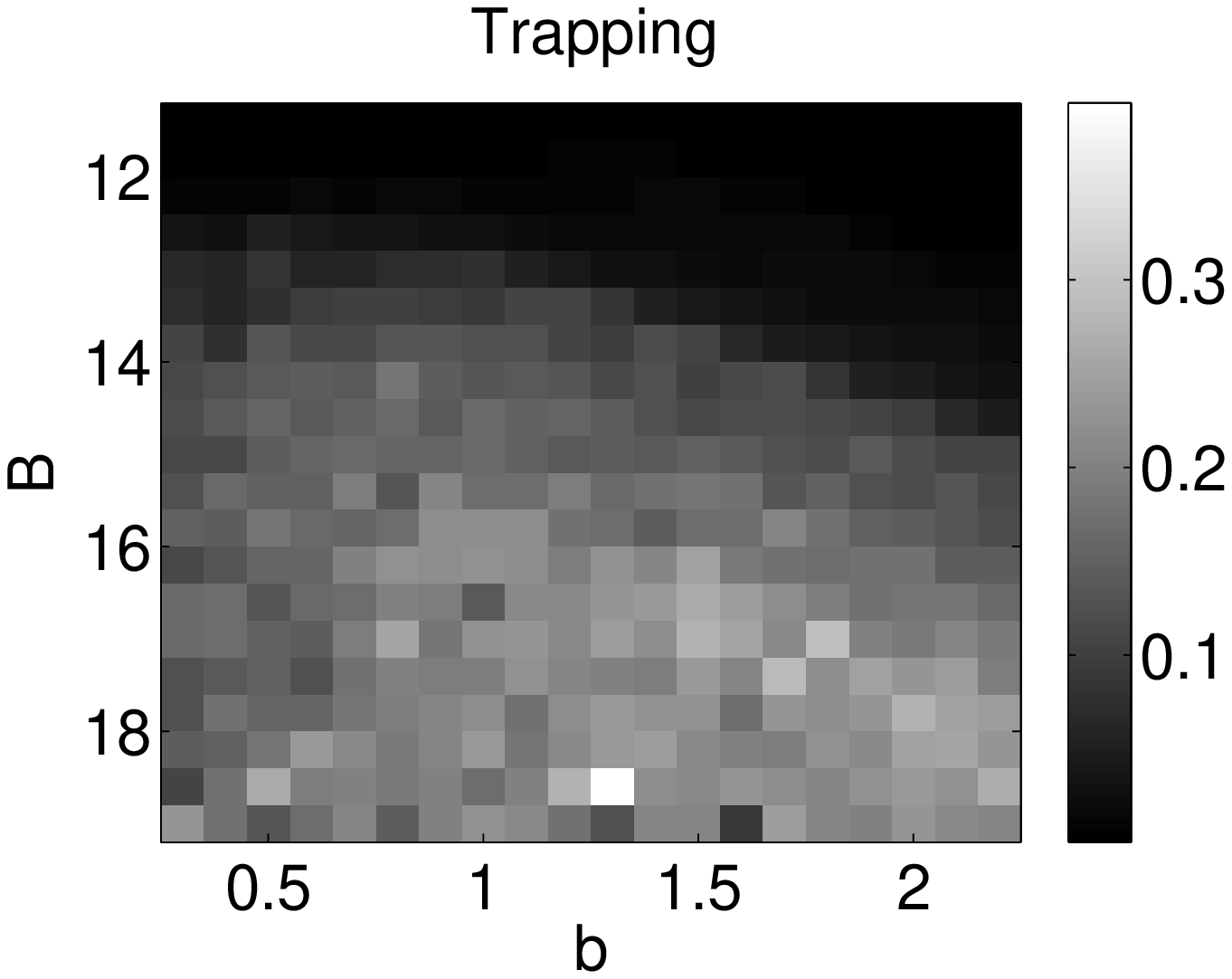}}} %
\subfigure[]{\scalebox{0.38}{\includegraphics{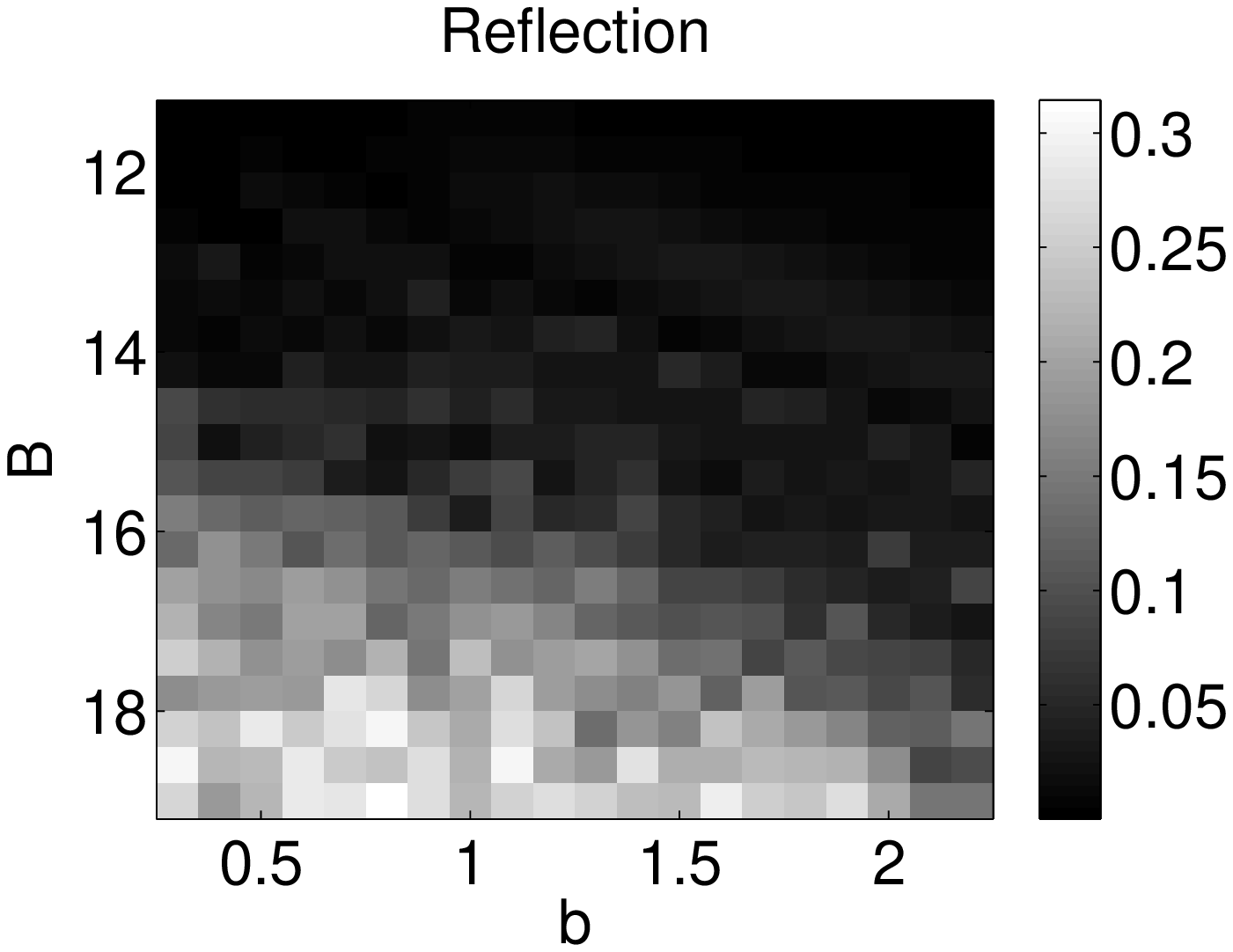}}} %
\subfigure[]{\scalebox{0.38}{\includegraphics{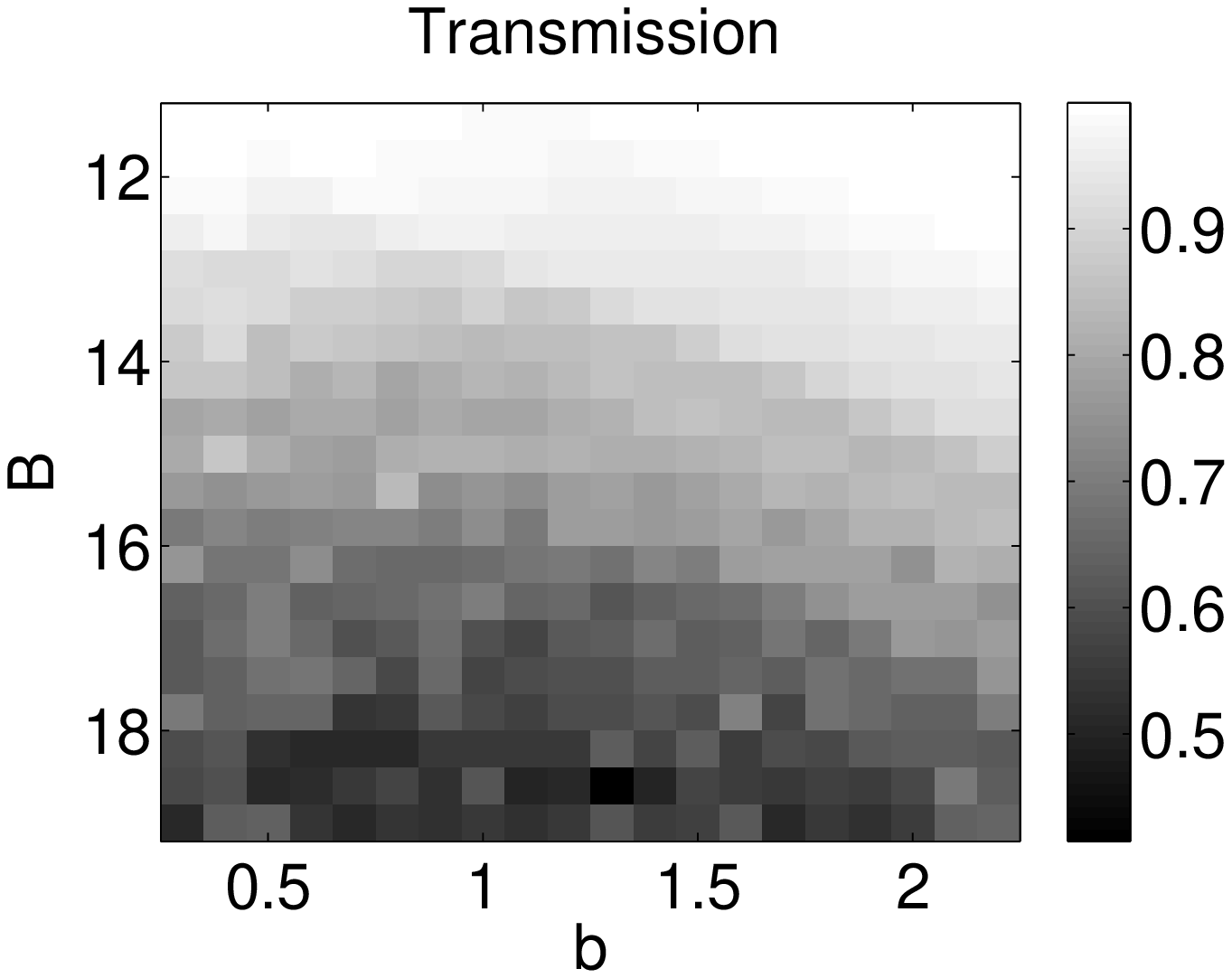}}}
\caption{Panels (a), (b), and (c) display, by means of gray-scale plots, the
trapped, reflected, and transmitted shares of the norm of the incident wave
packet, as functions of its amplitude ($B$) and inverse width ($b$), for
fixed velocity $c=20$ and width $\protect\epsilon =0.05$ of the nonlinear
scatterer, at the moment of time $t=0.5$. The trapped share is defined by
the norm confined in interval $-1<x<+1$. }
\label{fig6}
\end{figure}
\begin{figure}[tph]
\subfigure[]{\scalebox{0.4}{\includegraphics{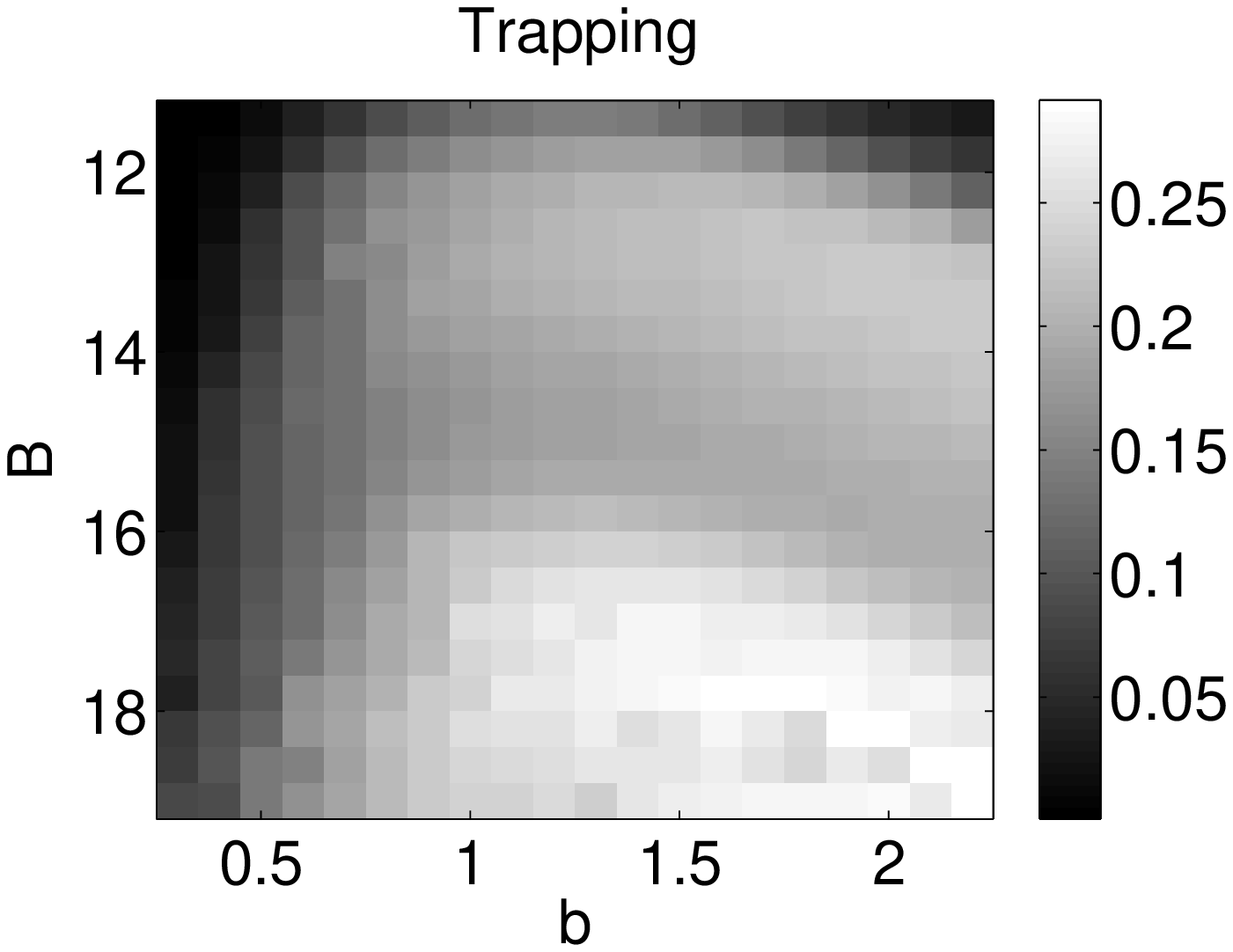}}}\subfigure[]{%
\scalebox{0.4}{\includegraphics{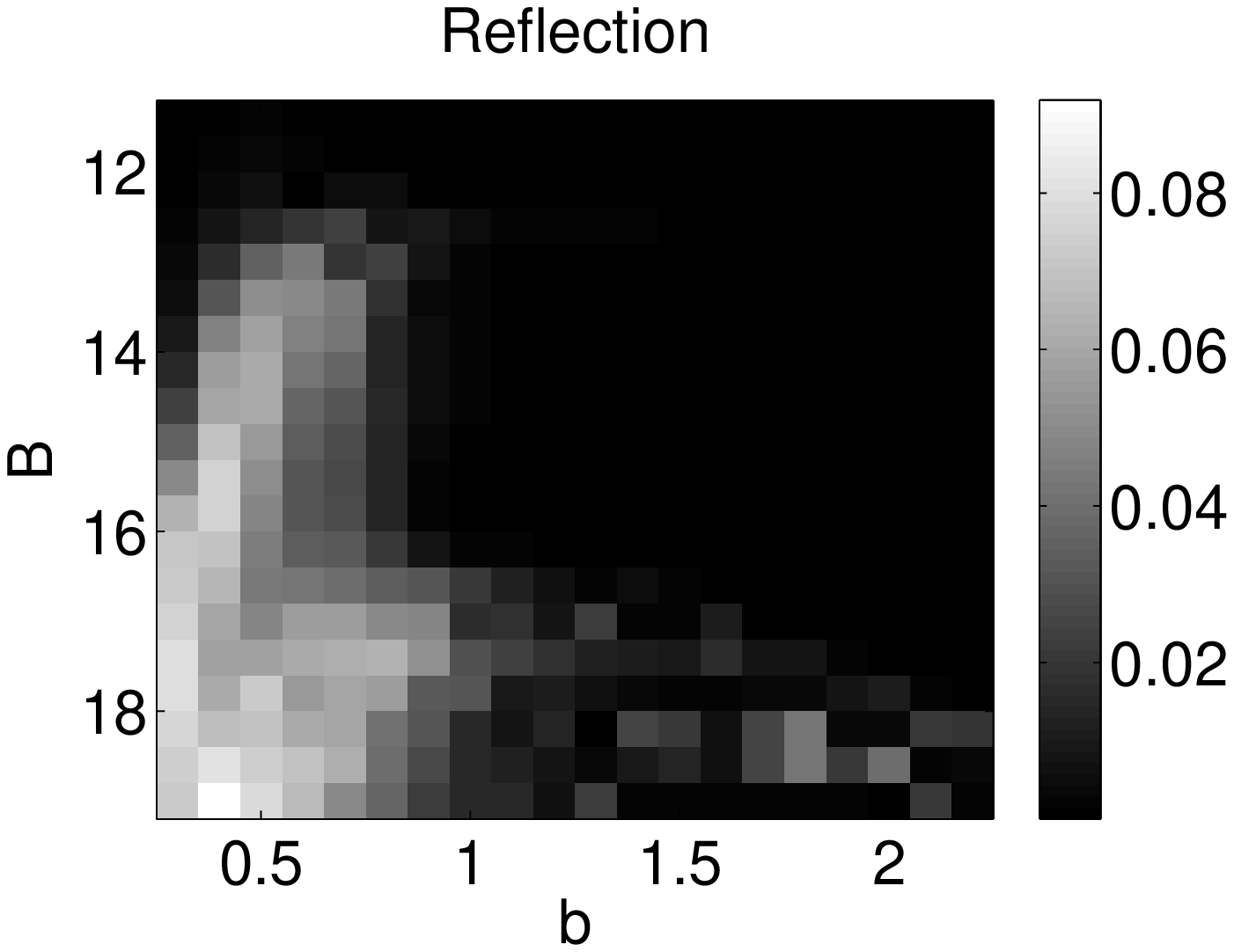}}} \subfigure[]{\scalebox{0.4}{%
\includegraphics{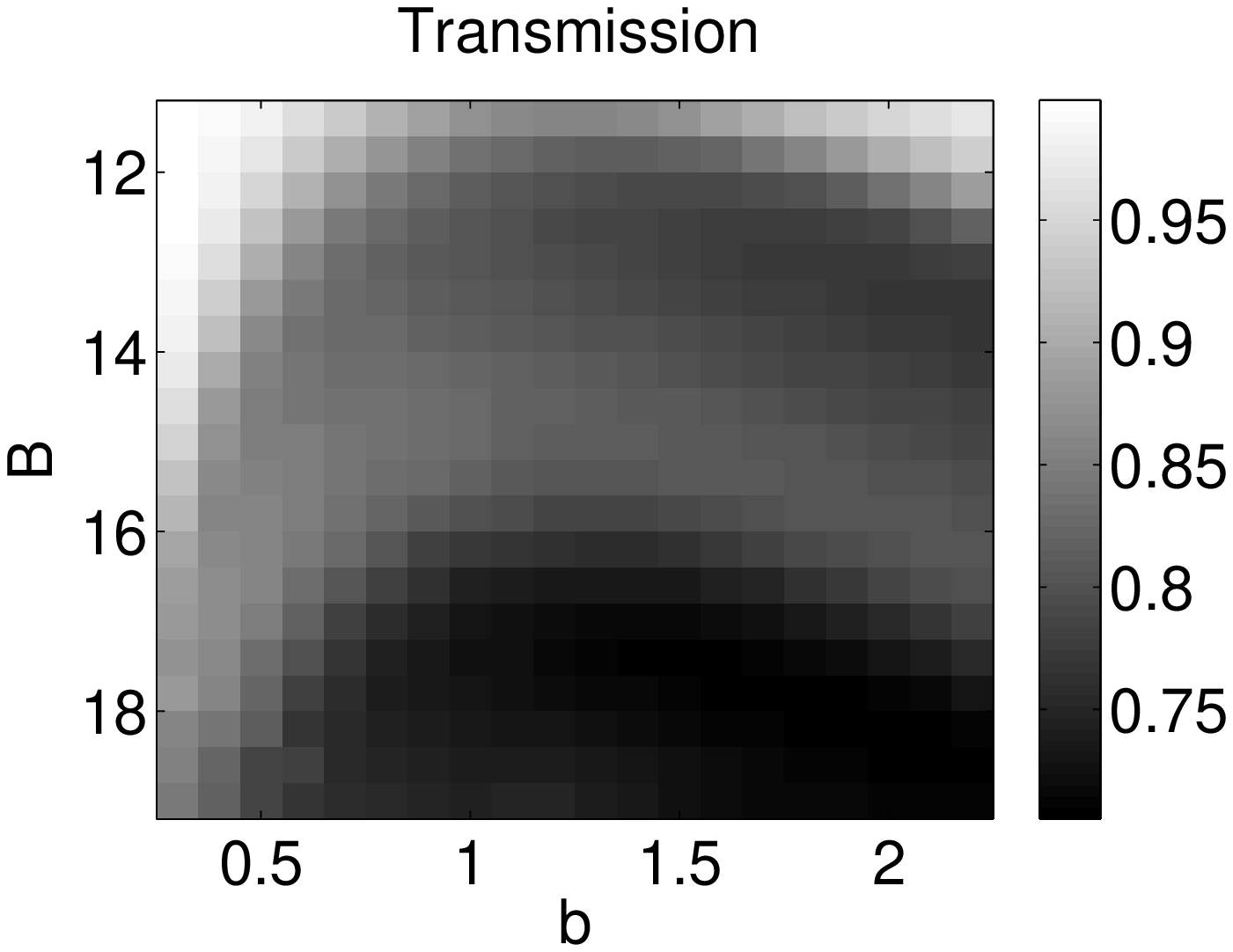}}}
\caption{The same as in Fig. \protect\ref{fig6}, but but for a broader
scatterer, with $\protect\epsilon =0.10$.}
\label{fig7}
\end{figure}

Further, the comparison of Figs. \ref{fig6} and \ref{fig7} shows that the
increase of the width leads to a gradual change of the scattering picture.
It is quite natural that the wider scatterer (Fig. \ref{fig7}) provides
weaker reflection but stronger trapping of the incident waves. Generally,
the solution of the scattering problem is more sensitive to the
width $\epsilon $
than the results reported above for the transfer problem, as in the latter
case the wave packet had a chance to adjust itself to the particular shape
of the nonlinear scatterer.

\section{Conclusion}

In this work, we have considered the 1D Schr\"{o}dinger model with the
tightly localized nonlinearity, to study the following dynamical problems:
the existence and stability of standing waves
 trapped by this nonlinear potential;
the controllable transfer of the localized mode by the moving nonlinear trap
(nonlinear tweezers); and the scattering of coherent wave packets on the
stationary localized nonlinearity. By means of systematic simulations, the
border of the effective stability has been identified for the transfer
problem. Relative shares of the norm (power) of the trapped, reflected, and
transmitted waves were found, as functions of the amplitude and width of the
incident pulse, for the scattering problem. The outcome of the transfer is
less sensitive to particular characteristics of the nonlinear potential trap
(such as its width) than the results of the solution of the scattering
problem. The quasi-Airy stationary modes, dragged by the nonlinear trap
moving at a constant acceleration, were briefly considered too, and were
found to be progressively less robust for beams with the increasing
amplitude. Some results were obtained in the approximate analytical form,
such as the adiabatic approximation for the slowly dragged localized mode,
and a qualitative explanation of the quasi-parabolic shape of the stability
border.

This work may be naturally extended in other directions. It is especially
interesting to consider the two-dimensional version of the transfer problem,
using the corresponding localized nonlinear trap in the form of a circle
\cite{Sakaguchi2,JMO} and setting it in motion. A challenging issue in this
case is the competition between the trapping of the two-dimensional wave
packet, and its propensity to the intrinsic collapse. It may be also
interesting to consider the scattering problem in one dimension in the case
of a localized \textit{repulsive} nonlinearity. Such studies will be
reported in future publications.

\section*{Acknowledgment}

B.A.M. appreciates hospitality of Institut de Ciencies Fotoniques at
Castelldefels (Barcelona). The work of D.J.F. was partially supported by the
Special Account for Research Grants of the University of Athens. P.G.K.
gratefully acknowledges support from the National Science Foundation through
grants DMS-0806762 and CMMI-1000337, as well as from the Alexander von
Humboldt Foundation and the Alexander S. Onassis Public Benefit Foundation.

\end{document}